\documentclass[12pt]{article} 
\usepackage{cite} 
\usepackage{hyperref}  
\begin{document} 

\begin{center} 
	 
{\large\bf Quantum mechanics of round magnetic electron lenses with Glaser and power law 
models of $B(z)$} 

\bigskip 
	 
Sameen Ahmed Khan$^{a,}$\footnote{Corresponding author \\    
{\em Email}:~\url{rohelakhan@yahoo.com}, \url{https://orcid.org/0000-0003-1264-2302}},  
Ramaswamy Jagannathan$^{b,}$\footnote{Retired Faculty (Theoretical Physics) \\   
{\em E-mail}:~\url{jagan@imsc.res.in}, \url{https://orcid.org/0000-0003-2968-2044}} 

\medskip 

$^a${\em Department of Mathematics and Sciences, \\ 
College of Arts and Applied Sciences (CAAS), Dhofar University, \\ 
Post Box No. 2509, Postal Code:~211, Salalah, Sultanate of Oman}   

\smallskip 

$^b${\em The Institute of Mathematical Sciences, \\ 
4th Cross Street, Central Institutes of Technology (CIT) Campus, \\ 
Tharamani, Chennai 600113, India}   

\bigskip 

\begin{abstract} 
Scalar theory of quantum electron beam optics, at the single-particle level, derived from the 
Dirac equation using a Foldy-Wouthuysen-like transformation technique is considered.  Round 
magnetic electron lenses with Glaser and power law models for the axial magnetic field $B(z)$ 
are studied.  Paraxial quantum propagator for the Glaser model lens is obtained in terms of the 
well known fundamental solutions of its paraxial equation of motion.  In the case of lenses with 
the power law model for $B(z)$ the well known fundamental solutions of the paraxial equations, 
obtained by solving the differential equation, are constructed using the Peano-Baker series also.  Quantum mechanics of aberrations is discussed briefly.  Role of quantum uncertainties in 
aberrations, and in the nonlinear part of the equations of motion for a nonparaxial beam, is 
pointed out.  The main purpose of this article is to understand the quantum mechanics of electron 
beam optics though the influence of quantum effects on the optics of present-day electron beam 
devices might be negligible.       
\end{abstract} 

\end{center}

\medskip 


\section{Introduction}  
In classical electron beam optics the trajectory of an electron moving in the electromagnetic 
field of any beam optical system, like an electron lens, is studied using classical mechanics 
and this study forms the basis for the design and operation of such beam optical devices.  
Theoretical, practical, and historical aspects of classical mechanics of electron beam optics 
have been presented extensively, with detailed references to the literature, in the first two 
volumes of the encyclopedic text book of Hawkes and Kasper \cite{Hawkes1989a, Hawkes1989b} and 
the third volume \cite{Hawkes1994} presents the wave mechanics of electron beam optics, or 
essentially, the quantum mechanics of electron optical imaging (for classical electron beam 
optics see also {\em e.g.}, \cite{ElKareh1970, Wollnik1987, Szilagyi1988, Liebl2008, Orloff2009, Rose2012}).  In studying the quantum mechanics of imaging in electron microscopy mostly the nonrelativistic Schr\"{o}dinger equation is used, following the pioneering work of Glaser 
\cite{Glaser1952, Glaser1953, Glaser1956} (see \cite{Hawkes1989a, Hawkes1989b, Hawkes1994} for 
details).  In relativistic situations, the Schr\"{o}dinger equation with relativistically 
corrected mass, Klein-Gordon equation, or an approximate scalar wave equation derived from the 
Dirac equation are used (for details, see \cite{Hawkes1994, Ferwerda1986a, Ferwerda1986b, Groves2015,Lubk2018, Pozzi2016}).  Driven by the curiosity to find how classical mechanics is so successful in the design and operation of charged particle beam optical devices like electron microscopes and particle accelerators when the microscopic charged particles passing through the systems should be obeying quantum mechanics, a systematic study of quantum mechanics of electron 
beam optics, at the single-particle level, derived from the Dirac equation using a 
Foldy-Woutuysen-like transformation technique was initiated in \cite{Jagannathan1989} and the 
first quantum mechanical derivation of the classical Busch formula for the focal length of an 
axially symmetric magnetic electron lens, or round magnetic lens, was obtained.  Since the electron 
is a spin-$\frac{1}{2}$ particle, Dirac equation is the proper equation on which a general theory 
of quantum electron beam optics has to be based.  Of course, approximations applicable to cases 
like nonrelativistic situations and treatments ignoring spin can be derived from such a general theory.  With this in mind, following \cite{Jagannathan1989}, a formalism of quantum charged 
particle beam optics applicable to devices from low energy electron microscopes to high energy 
particle accelerators is being developed \cite{Jagannathan1990, Khan1995, Jagannathan1996, 
Conte1996, Khan1997, Jagannathan1999, Khan1999a, Khan1999b, Jagannathan2002, Khan2002a, Jagannathan2004, Khan2016, Khan2018a}. 

A consolidated account of the formalism of quantum mechanics of charged particle beam optics, at  
the single-particle level, has been presented in \cite{Jagannathan2019} where we have studied the  general theory of round magnetic lenses based on relativistic quantum mechanics using both the 
Klein-Gordan equation (ignoring spin of the particle) and the Dirac equation (taking into account 
the spin-$\frac{1}{2}$ nature of particles like electrons) without assuming any specific model for 
the magnetic field of the system.  We have shown, in general, how the relativistic quantum theory 
of any charged particle beam optical system can be approximated leading to the nonrelativistic 
quantum theory of the system.  Following \cite{Jagannathan2019}, we have studied in \cite{Khan2020} 
the quantum mechanics of bending of a nonrelativistic monoenergetic charged particle beam by a 
dipole magnet in the paraxial approximation.  In this article we shall study round magnetic lenses 
with Glaser and power law models for the axial magnetic field on the basis of the scalar quantum 
theory derived from the Dirac equation using the Foldy-Wouthuysen-like transformation technique.   

Let us consider a round magnetic lens with the $z$-axis as its straight optic axis.  The  system is formed by the axially symmetric magnetic field obtained from the vector potential 
(see {\em e.g.}, \cite{Hawkes1989a})     
\begin{equation} 
\vec{A}\left(\vec{r}_\perp,z\right) 
   = \left(-\frac{1}{2}y\Pi\left(\vec{r}_\perp,z\right),
     \frac{1}{2}x\Pi\left(\vec{r}_\perp,z\right),0\right), 
\label{vecA}
\end{equation}   
with 
\begin{eqnarray}
\Pi\left(\vec{r}_\perp,z\right) 
   & = & \sum_{n=0}^{\infty}\frac{1}{n!(n+1)!}
         \left(-\frac{r_\perp^2}{4}\right)^nB^{(2n)}(z)  \nonumber \\ 
   & = & B(z) - \frac{1}{8}B^{\prime\prime}(z)r_\perp^2 
              + \frac{1}{192}B^{(4)}(z)r_\perp^4 - \ldots, 
\end{eqnarray}
where $\vec{r}_\perp = (x,y)$, $r_\perp^2 = x^2+y^2$, and  
\begin{eqnarray} 
B^{(0)}(z) 
   & = & B(z),\quad B^\prime(z) = \frac{dB(z)}{dz}, \quad 
         B^{\prime\prime}(z) = \frac{d^2B(z)}{dz^2},  \nonumber \\ 
B^{\prime\prime\prime}(z) 
   & = & \frac{d^3B(z)}{dz^3}, \quad \ldots, \quad B^{(2n)}(z) = \frac{d^{2n}B(z)}{dz^{2n}}. 
\end{eqnarray}   
The corresponding magnetic field $\vec{B} = \vec{\nabla}\times\vec{A}$ has the components  
\begin{eqnarray} 
\vec{B}_\perp\left(\vec{r}_\perp,z\right)
   & = & -\frac{1}{2}\left(B^\prime(z) - \frac{1}{8}B^{\prime\prime\prime}(z)r_\perp^2 
         + \ldots\right)\vec{r}_\perp, \nonumber \\ 
B_z\left(\vec{r}_\perp,z\right)  
   & = & B(z) - \frac{1}{4}B^{\prime\prime}(z)r_\perp^2 + \frac{1}{64}B^{(4)}(z)r_\perp^4 
              - \ldots. 
\label{magfield}
\end{eqnarray}  
Thus, $B(z) = B_z(0,0,z)$ is the axial magnetic field of such systems.  For the classic Glaser 
model lens $B(z)$ is given by the Lorentzian curve 
\begin{equation}  
B(z) = \frac{B_0}{1 + (z/a)^2}, 
\label{Glaser}
\end{equation} 
where $B_0 = B(0)$ is the value of the field at the maximum of the bell-shaped distribution and 
$a$ is the half-width of the distribution.  Besides the Glaser model lens, we shall study the 
power law model lenses (see \cite{Hawkes1989a, Hawkes1989b, Hawkes1982, Gianola1952, Hansel1964, 
Alshwaikh1977, AlHilly1982, Mulvey1982, Lenc1992, Crewe2001, Hawkes2002, Liu2003, Crewe2003, 
Crewe2004, Alamir2003, Alamir2004, Alamir2005, Alamir2009a, Alamir2009b, Alamir2011}) for which 
\begin{equation} 
B(z) \propto z^n. 
\label{powerlaw} 
\end{equation}   
Optical properties of these lenses have been analysed, based on classical electron optics, very extensively using diverse analytical and numerical tools.  The purpose of this article is mainly 
to understand the quantum mechanics of such lenses though the quantum corrections to the classical results might be negligible.   

\section{Quantum electron beam optical Hamiltonian of a round magnetic lens}  
Let us consider a monoenergetic, quasiparaxial {\em i.e.}, almost paraxial, beam of electrons 
moving along the positive $z$-axis of a round lens comprising of a static magnetic field $\vec{B}(\vec{r})$ given in (\ref{magfield}).  In an ideal paraxial beam any electron will have 
$|\vec{p}_\perp| \approx 0$.  For such a paraxial beam we can take 
\begin{equation} 
\Pi\left(\vec{r}_\perp,z\right) \approx B(z)  
\end{equation} 
in (\ref{vecA}).  In a quasiparaxial beam, slightly deviating from the paraxial condition, we 
could have $|\vec{p}_\perp| \ll p_z$.  For such a quasiparaxial beam we can take 
\begin{equation}
\Pi\left(\vec{r}_\perp,z\right) \approx B(z) - \frac{1}{8}B^{\prime\prime}(z)r_\perp^2.  
\label{nonparax}
\end{equation}
In the paraxial case only the lowest order terms in $\vec{r}_\perp$ would contribute to the 
effective field felt by the beam electrons moving close to the optic axis.  Let us denote the 
rest mass of the electron by $m$ and its charge, $-e$, by $q$.  The Dirac equation governing 
the quantum dynamics of the electron, ignoring its anomalous magnetic moment, is 
\begin{equation}
i\hbar\frac{\partial\underline{\Psi}\left(\vec{r},t\right)}{\partial t} 
   = \hat{H}\underline{\Psi}\left(\vec{r},t\right),  \quad    
\underline{\Psi}\left(\vec{r},t\right) 
   = \left(\begin{array}{c}
           \Psi_1(\vec{r},t) \\ 
           \Psi_2(\vec{r},t) \\ 
   	       \Psi_3(\vec{r},t) \\
   	       \Psi_4(\vec{r},t) 
           \end{array}\right),   
\label{DiracEq}
\end{equation} 
where $\underline{\Psi}\left(\vec{r},t\right)$ is the $4$-component Dirac spinor wave function associated with the electron and $\hat{H}$ is the Dirac Hamiltonian given by 
\begin{eqnarray} 
\hat{H}
   & = & \left(\begin{array}{cccc} 
               mc^2 & 0 & c\hat{\pi}_z & c\left(\hat{\pi}_x-i\hat{\pi}_y\right) \\ 
               0  & mc^2 & c\left(\hat{\pi}_x+i\hat{\pi}_y\right) & -c\hat{\pi}_z \\ 
               c\hat{\pi}_z & c\left(\hat{\pi}_x-i\hat{\pi}_y\right) & -mc^2 & 0 \\ 
               c\left(\hat{\pi}_x+i\hat{\pi}_y\right) & -c\hat{\pi}_z & 0 & -mc^2 
               \end{array}\right)  \nonumber \\    
   & = & mc^2\beta + c\vec{\alpha}\cdot\vec{\hat{\pi}},   
\label{DiracH}
\end{eqnarray} 
with 
\begin{eqnarray} 
\vec{\hat{\pi}} 
   & = & \vec{\hat{p}} - q\vec{A} = -i\hbar\vec{\nabla} - q\vec{A}, \\ 
\beta 
   & = & \left(\begin{array}{cr} 
               \mathbf{1} & \mathbf{0} \\ 
               \mathbf{0} & -\mathbf{1} \\   
               \end{array}\right), \quad  
\vec{\alpha} 
   = \left(\begin{array}{cc} 
      	   \mathbf{0} & \vec{\sigma} \\ 
	       \vec{\sigma} & \mathbf{0} \\   
           \end{array}\right),  \nonumber \\  
\sigma_x 
   & = & \left(\begin{array}{cc}
	           0 & 1 \\ 1 & 0
               \end{array}\right),  \quad  
\sigma_y 
   = \left(\begin{array}{lr}
	       0 & -i \\ i & 0 
           \end{array}\right),  \quad  
\sigma_z 
  = \left(\begin{array}{lr}
	      1 & 0 \\ 0 & -1
          \end{array}\right),  
\end{eqnarray} 
in which 
\begin{equation}
\mathbf{1} = \left(\begin{array}{cc} 
          	       1 & 0 \\ 0 & 1 \\   
                   \end{array}\right), \quad  
\mathbf{0} = \left(\begin{array}{cc} 
	               0 & 0 \\ 0 & 0 \\   
                   \end{array}\right).  	
\end{equation}  
The three Pauli matrices $\vec{\sigma}$ and the four Dirac matrices 
$\left(\vec{\alpha},\beta\right)$ satisfy the relations,   
\begin{eqnarray}
\sigma_j\sigma_k 
   & = & -\sigma_k\sigma_j, \ \ \mbox{for}\ j \neq k, \ \ j,k = x,y,z, \quad 
\sigma_j^2 
     = \mathbf{1},\ \  j = x,y,z, \nonumber \\ 
\alpha_j\alpha_k 
   & = & -\alpha_k\alpha_j, \ \ \mbox{for}\ j \neq k, \ \ j,k = x,y,z, \quad 
\alpha_j^2 
     = I, \ \ j = x,y,z, \nonumber \\ 
\beta\alpha_j 
   & = & -\alpha_j\beta, \ \  j = x,y,z, \quad \beta^2 = I,    
\label{DiracAlgebra}
\end{eqnarray} 
where  
\begin{equation} 
I = \left(\begin{array}{cc} 
	      \mathbf{1} & \mathbf{0} \\ 
	      \mathbf{0} & \mathbf{1} \\   
          \end{array}\right).  	
\end{equation}
When an electron is associated with the $4$-component wave function  $\underline{\Psi}\left(\vec{r},t\right)$ the probability for it to be found at the position 
$\vec{r}$ at time $t$ is given by 
\begin{equation} 
\underline{\Psi}\left(\vec{r},t\right)^\dagger\underline{\Psi}\left(\vec{r},t\right) 
   = \sum_{j=1}^{4}\Psi_j^*\left(\vec{r},t\right)\Psi_j\left(\vec{r},t\right).   
\label{probatt}
\end{equation} 
Hence $\underline{\Psi}\left(\vec{r},t\right)$ is normalized as 
\begin{equation}
\int d^3r\,\underline{\Psi}\left(\vec{r},t\right)^\dagger
           \underline{\Psi}\left(\vec{r},t\right) = 1, 
\label{normalzn}
\end{equation} 
so that the probability of finding the electron somewhere in space is one at any time $t$, where 
$\int d^3r$ stands for 
$\int_{-\infty}^{\infty}\int_{-\infty}^{\infty}\int_{-\infty}^{\infty}dxdydz$.  Since the Dirac Hamiltonian in (\ref{DiracH}) is Hermitian the time evolution of $\underline{\Psi}\left(\vec{r},t\right)$ according to (\ref{DiracEq}) is unitary and the 
normalization (\ref{normalzn}) is preserved in time.
 
We are interested in studying the evolution of the beam along the optic axis {\em i.e.}, $z$-axis.  Since the beam is monoenergetic and the field is time-independent, with $\vec{r} = (x,y,z)$ 
written as $\left(\vec{r}_\perp,z\right)$, we can take 
\begin{equation}
\underline{\Psi}(\vec{r}_\perp,z,t) 
   = \exp{(-iEt/\hbar)}\underline{\psi}(\vec{r}_\perp,z)  
   = \exp{(-iEt/\hbar)}\left(\begin{array}{c}
                                 \psi_1(\vec{r}_\perp,z)\\ 
                                 \psi_2(\vec{r}_\perp,z) \\ 
                                 \psi_3(\vec{r}_\perp,z) \\
                                 \psi_4(\vec{r}_\perp,z) 
                                 \end{array}\right),   
\label{Psi}
\end{equation} 
where $E = \sqrt{m^2c^4+c^2|\vec{p}_0|^2}$ is the conserved total energy of the electron with 
$|\vec{p}_0| = p_0$ as the design momentum.  Note that $E$ is positive for us. Let us write 
$\vec{p}_0 = (\vec{p}_{0\perp},p_{0z})$.    Since the beam is quasiparaxial and is moving along 
the positive $z$-axis, $p_{0z} > 0$, $p_{0z} \approx p_0$ and 
$\left|\vec{p}_{0\perp}\right| \ll p_0$.  With the choice of $\underline{\Psi}$ as in 
(\ref{Psi}) the time-dependent Dirac equation (\ref{DiracEq}) becomes 
\begin{equation}
\hat{H}\underline{\psi} = E\underline{\psi},    
\label{psi} 
\end{equation} 
the time-independent Dirac equation.  With the beam of electrons entering the system through 
one side and leaving it through the opposite side, we are dealing with the scattering states of 
the system, not the bound stationary states.  We have to study the $z$-evolution of the wave 
function of the beam propagating through the system in order to understand the optical 
properties of the system.  To this end we proceed as follows.    

Multiplying both sides of (\ref{psi}) from left by ${\alpha_z}/c$ and rearranging the terms we 
get the $z$-evolution equation for $\underline{\psi}$ given by   
\begin{equation}
i\hbar\frac{\partial\underline{\psi}(\vec{r}_\perp,z)}{\partial z} 
   = \hat{\mathcal{H}}\underline{\psi}(\vec{r}_\perp,z),  \quad 
\hat{\mathcal{H}} 
   = -p_0\beta\chi\alpha_z - qA_zI + \alpha_z\vec{\alpha}_\perp\cdot\vec{\hat{\pi}}_\perp, 
\label{zevoln}
\end{equation} 
with  
\begin{equation} 
p_0 = \frac{1}{c}\sqrt{E^2 - m^2c^4}, \quad  
\chi = \left(\begin{array}{cr} 
             \xi\mathbf{1} & \mathbf{0} \\ 
             \mathbf{0} & -\xi^{-1}\mathbf{1} \\   
             \end{array}\right), \quad 
\xi = \sqrt{\frac{E + mc^2}{E - mc^2}}\,,  
\end{equation} 
where the square roots are taken to be positive.  Now, let 
\begin{equation} 
\underline{\psi}^{\prime} 
   = M\underline{\psi},  \qquad   
M  = \frac{1}{\sqrt{2}}\left(I + \chi\alpha_z\right), \quad  
M^{-1} = \frac{1}{\sqrt{2}}\left(I - \chi\alpha_z\right).  
\end{equation} 
Then, we get 
\begin{eqnarray} 
i\hbar\frac{\partial\underline{\psi}^\prime}{\partial z} 
   & = & \hat{\mathcal{H}}^\prime\underline{\psi}^\prime, \quad 
\hat{\mathcal{H}}^\prime 
    =  M\hat{\mathcal{H}}M^{-1} 
    = -p_0\beta + \hat{\cal E} + \hat{\cal O}, \nonumber \\  
\hat{\cal E} 
   & = & -qA_z\left(\begin{array}{cr} 
                   \mathbf{1} & \mathbf{0} \\ \mathbf{0}& -\mathbf{1} \\   
                   \end{array}\right),  \quad 
\hat{\cal O} 
    = \left(\begin{array}{cc} 
               \mathbf{0} & \xi\vec{\alpha}_\perp\cdot\vec{\hat{\pi}}_\perp \\ 
              -\xi^{-1}\vec{\alpha}_\perp\cdot\vec{\hat{\pi}}_\perp & \mathbf{0} \\   
               \end{array}\right).  
\label{Hprime} 
\end{eqnarray} 
The Dirac Hamiltonian $\hat{H}$ in (\ref{DiracH}) is a sum of two parts, the diagonal part 
$mc^2\beta$ and the off-diagonal part $c\vec{\alpha}\cdot\vec{\hat{\pi}}$ which contains 
$2\times 2$ matrix blocks only along the off-diagonal and anticommutes with $\beta$.  If the 
system has an electric field also the scalar potential will contribute another diagonal term 
to the Hamiltonian.  In general, in the Dirac Hamiltonian the diagonal part which commutes 
with $\beta$ and does not couple the upper and lower pairs of the four-component Dirac wave 
function is called even and the off-diagonal part which anticommutes with $\beta$ and couples 
the upper and lower pairs of the Dirac wave function is called odd.  This structure of 
$\hat{H}$ helps expand it as a series of terms, with $1/mc^2$ as the expansion parameter, 
that correspond to the nonrelativistic approximation and relativistic corrections of various 
orders.  Actually, the dimensionless expansion parameter is $c\hat{p}/mc^2$, as in the expansion 
of the free particle energy 
\begin{eqnarray}
	E & = & \sqrt{m^2c^4 + c^2p^2} 
	= mc^2\sqrt{1 + \left(\frac{cp}{mc^2}\right)^2}  \nonumber \\ 
	& = & mc^2\left[1 + \frac{1}{2}\left(\frac{cp}{mc^2}\right)^2 
	- \frac{1}{8}\left(\frac{cp}{mc^2}\right)^4 + \ldots \right] 
	\approx mc^2 + \frac{p^2}{2m} - \frac{p^4}{8m^3c^2}\,,   
\end{eqnarray}
but, for book keeping purposes one usually regards $1/mc^2$ as the expansion parameter.  The desired 
expansion of $\hat{H}$ as a series of approximations, up to any desired order, is obtained via the Foldy-Wouthuysen transformation technique which consists of a series of transformations till the desired order of accuracy is reached (see {\em e.g.}, \cite{Bjorken1994}).  A similar structure is 
seen in $\hat{\mathcal{H}}^\prime$ (\ref{Hprime}) in which the diagonal part, 
$-p_0\beta +\hat{\cal E}$, commutes with $\beta$, and the off-diagonal part, 
$\hat{\cal O}$, anticommutes with $\beta$.  This helps us expand $\hat{\mathcal{H}}^\prime$ 
(or $\hat{\mathcal{H}} = M^{-1}\hat{\mathcal{H}}^\prime M$) as a series of approximations, 
with $1/p_0$ as the expansion parameter, starting with the paraxial approximation and followed by nonparaxial approximations.  Here again, the dimensionless expansion parameter is $\hat{p}_\perp/p_0$, like in 
the expression 
\begin{eqnarray} 
	p_z & = & \sqrt{p_0^2 - p_\perp^2} 
	= p_0\sqrt{1 - \left(\frac{p_\perp}{p_0}\right)^2}  \nonumber \\ 
	& = & p_0\left[1 - \frac{1}{2}\left(\frac{p_\perp}{p_0}\right)^2 
	- \frac{1}{8}\left(\frac{p_\perp}{p_0}\right)^4 - \ldots \right] 
	\approx p_0 -\frac{p_\perp^2}{2p_0} - \frac{p_\perp^4}{8p_0^3}\,, 
\end{eqnarray}
for a free particle, and one uses $1/p_0$ as the expansion parameter for book keeping purposes.  
To get the desired result we shall be using a Foldy-Wouthuysen-like (FW-like) transformation technique.    

The first FW-like transformation is 
\begin{equation}
\underline{\psi}^{(1)} = \exp{\left(i\hat{S}_1\right)}\underline{\psi}^\prime, \qquad 
\hat{S}_1 = \frac{i}{2p_0}\beta\hat{\cal O}.  
\end{equation} 
This leads to the $z$-evolution equation for $\underline{\psi}^{(1)}$ given by  
\begin{equation}
i\hbar\frac{\partial\underline{\psi}^{(1)}}{\partial z} 
   = \hat{\mathcal{H}}^{(1)}\underline{\psi}^{(1)}, 
\end{equation} 
where 
\begin{eqnarray} 
\hat{\mathcal{H}}^{(1)} 
   & = &  \exp{\left(-\beta\hat{\cal O}/2p_0\right)}\hat{\mathcal{H}}^\prime
          \exp{\left(\beta\hat{\cal O}/2p_0\right)}  \nonumber \\  
   &   & - i\hbar\exp{\left(-\beta\hat{\cal O}/2p_0\right)}
          \frac{\partial}{\partial z}\left(\exp{\left(\beta\hat{\cal O}/2p_0\right)}\right) 
          \nonumber \\ 
   & \approx & -p_0\beta + \hat{\cal E}^{(1)} + \hat{\cal O}^{(1)},  \nonumber \\ 
\hat{\cal E}^{(1)} 
   & = &  \hat{\cal E} - \frac{1}{2p_0}\beta\hat{\cal O}^2  
        - \frac{1}{8p_0^2}\left[\hat{\cal O},\left(\left 
                [\hat{\cal O},\hat{\cal E}\right]
        + i\hbar\frac{\partial\hat{\cal O}}{\partial z}\right)\right] 
        - \frac{1}{8p_0^3}\beta \hat{\cal O}^4,  \nonumber \\ 
\hat{\cal O}^{(1)} 
   & = & - \frac{1}{2p_0}\beta\left(\left[\hat{\cal O},\hat{\cal E}\right]
         + i\hbar\frac{\partial\hat{\cal O}}{\partial z}\right) 
         - \frac{1}{3p_0^2}\hat{\cal O}^3,   
\label{FirstFWT} 
\end{eqnarray} 
in which $\hat{\cal E}^{(1)}$ is even, $\hat{\cal O}^{(1)}$ is odd, and  
$\left[\hat{A},\hat{B}\right] = \hat{A}\hat{B} - \hat{B}\hat{A}$, the commutator of $\hat{A}$ 
and $\hat{B}$.  In deriving $\hat{\mathcal{H}}^{(1)}$ we have used the relations   
\begin{eqnarray}
\exp{(-\hat{A})}B\exp{(\hat{A})}  
   & = & \hat{B} - \left[\hat{A},\hat{B}\right]  
                 + \frac{1}{2!}\left[\hat{A},\left[\hat{A},\hat{B}\right]\right]  \nonumber \\ 
   &   & \quad   - \frac{1}{3!}\left[\hat{A},\left[\hat{A},\left[\hat{A},\hat{B}\right]
                   \right]\right] + \ldots,  \\ 
\exp{(-\hat{A}(z))}\frac{\partial}{\partial z}\left(\exp{(\hat{A}(z))}\right)   
   & = & \frac{\partial\hat{A}(z)}{\partial z}
                 - \frac{1}{2!}\left[\hat{A}(z),\frac{\partial\hat{A}(z)}{\partial z}\right] 
                   \nonumber \\ 
   &   & \qquad\quad + \frac{1}{3!}\left[\hat{A}(z),\left[\hat{A}(z),\frac{\partial\hat{A}(z)}
                {\partial z}\right]\right] + \ldots, 
\end{eqnarray} 
and only terms of order up to $1/p_0^3$ have been retained 
(see \cite{Wilcox1967, Jagannathan1996, Jagannathan2019} for more details).  Now, if we make 
a second FW-like transformation following the same recipe as for the first one, 
\begin{equation}
\underline{\psi}^{(2)} = \exp{\left(i\hat{S}_2\right)}\underline{\psi}^{(1)}, \qquad 
\hat{S}_2 = \frac{i}{2p_0}\beta\hat{\cal O}^{(1)},    
\end{equation} 
we get 
\begin{eqnarray}
i\hbar\frac{\partial\underline{\psi}^{(2)}}{\partial z} 
   & = & \hat{\mathcal{H}}^{(2)}\underline{\psi}^{(2)},  \quad 
\hat{\mathcal{H}}^{(2)} 
    \approx -p_0\beta + \hat{\cal E}^{(2)} + \hat{\cal O}^{(2)},  \nonumber \\ 
\hat{\cal E}^{(2)} 
   & = &  \hat{\cal E}^{(1)} - \frac{1}{2p_0}\beta\left(\hat{\cal O}^{(1)}\right)^2  
        - \frac{1}{8p_0^2}\left[\hat{\cal O}^{(1)},\left(\left
         [\hat{\cal O}^{(1)},\hat{\cal E}^{(1)}\right]
        + i\hbar\frac{\partial\hat{\cal O}^{(1)}}{\partial z}\right)\right]  \nonumber \\   
   &   & - \frac{1}{8p_0^3}\beta\left(\hat{\cal O}^{(1)}\right)^4,  \nonumber \\ 
\hat{\cal O}^{(2)} 
   & = & - \frac{1}{2p_0}\beta\left(\left[\hat{\cal O}^{(1)},\hat{\cal E}^{(1)}\right]
         + i\hbar\frac{\partial\hat{\cal O}^{(1)}}{\partial z}\right)  
         - \frac{1}{3p_0^2}\left(\hat{\cal O}^{(1)}\right)^3,     
\end{eqnarray} 
where the expressions for $\hat{\cal E}^{(1)}$ and $\hat{\cal O}^{(1)}$ are to be inserted 
from (\ref{FirstFWT}).  We can continue this series of FW-like transformations till we achieve 
the desired accuracy.  To this end, observe that $\hat{\cal O}$ in $\hat{\mathcal{H}}^\prime$ 
is of order zero in $1/p_0$, $\hat{\cal O}^{(1)}$ in $\hat{\mathcal{H}}^{(1)}$ is of first order 
in $1/p_0$, and $\hat{\cal O}^{(2)}$ in $\hat{\mathcal{H}}^{(2)}$ is of second order in $1/p_0$.  
This shows that with each transformation the odd operator in the Hamiltonian becomes weaker and 
we can stop the transformation process when the odd operator becomes weak to any desired order.  Stopping with the third FW-like transformation we get 
\begin{equation}
i\hbar\frac{\partial\underline{\psi}^{(3)}}{\partial z} 
   = \hat{\mathcal{H}}^{(3)}\underline{\psi}^{(3)},  \quad  
\hat{\mathcal{H}}^{(3)} 
   \approx -p_0\beta + \hat{\cal E}^{(3)}, 
\label{after3rdFW}
\end{equation} 
where we have dropped the odd term $\hat{\cal O}^{(3)}$ as negligible for our purpose.  
Further, note that for a quasiparaxial beam moving in the positive $z$ direction 
$\underline{\psi}$, or $\underline{\psi}^{(3)}$, will be like $\sim \exp{(ip_0z/\hbar)}$ such 
that  $i\hbar\partial\underline{\psi}^{(3)}/{\partial z} \sim -p_0\underline{\psi}$.  This 
would require the lower pair of components of $\underline{\psi}^{(3)}$ to be very small 
compared to the upper pair of components.  In view of this, $\hat{\mathcal{H}}^{(3)}$ can be approximated further by dropping $\beta$s from it, or replacing $\beta$s in it by $2\times 2$ 
identity matrices (see \cite{Jagannathan1996, Jagannathan2019} for more details).  

Let us now retrace the transformations we have made so that we go back to the original Dirac 
$\underline{\psi}$ in (\ref{zevoln}):  
\begin{equation}
\underline{\psi} = M^{-1}\exp{\left(\beta\hat{\cal O}/2p_0\right)}
                   \exp{\left(\beta\hat{\cal O}^{(1)}/2p_0\right)}
                   \exp{\left(\beta\hat{\cal O}^{(2)}/2p_0\right)}\underline{\psi}^{(3)}.  
\label{inverseFWT}
\end{equation} 
This inverse transformation will not take us to the original Hamiltonian $\hat{\mathcal{H}}$ in (\ref{zevoln}) since it has been expanded in a series and truncated.  Note that the Hamiltonian $\hat{\mathcal{H}}$ is not Hermitian.  We are considering the beam to be quasiparaxial {\em i.e.},  slightly deviating from the paraxial condition, and hence we can choose $\vec{A}$ to be given by (\ref{vecA}) and (\ref{nonparax}).  Then, carrying out the transformation (\ref{inverseFWT}), and keeping only terms of order up to $1/p_0^3$, and dropping the non-Hermitian terms, we get 
\begin{equation}
i\hbar\frac{\partial\underline{\psi}\left(\vec{r}_\perp,z\right)}{\partial z} 
   = \hat{\mathcal{H}}_o\underline{\psi}\left(\vec{r}_\perp,z\right),  
\label{QBOevoln}  
\end{equation}  
where   
\begin{eqnarray} 
\hat{\mathcal{H}}_o
   & = & \hat{\mathcal{H}}_{o,p} + \hat{\mathcal{H}}^\prime_o 
        +\hat{\mathcal{H}}_o^{(\hbar)} + \hat{\mathcal{H}}_{o,\sigma}^{(\hbar)},  
\label{DiracHo}\\ 
\hat{\mathcal{H}}_{o,p}  
   & = & -p_0 + \frac{1}{2p_0}\hat{p}_\perp^2  
         + \frac{1}{2}p_0\alpha^2(z)r_\perp^2 - \alpha(z)\hat{L}_z, 
\label{Hop}  \\ 
\hat{\mathcal{H}}^\prime_o 
   & = &   \frac{1}{8p_0^3}\hat{p}_\perp^4 
         - \frac{\alpha(z)}{2p_0^2}\hat{p}_\perp^2\hat{L}_z  
         - \frac{\alpha^2(z)}{8p_0}\left(\vec{r}_\perp\cdot\vec{\hat{p}}_\perp 
             + \vec{\hat{p}}_\perp\cdot\vec{r}_\perp\right)^2 \nonumber \\ 
   &   & + \frac{3\alpha^2(z)}{8p_0}\left(r_\perp^2\hat{p}_\perp^2 
         + \hat{p}_\perp^2r_\perp^2\right)  
         + \frac{1}{8}\left(\alpha^{\prime\prime}(z)-4\alpha^3(z)\right)
           \hat{L}_zr_\perp^2  \nonumber \\ 
   &   & + \frac{p_0}{8}\left(\alpha^4(z) - \alpha(z)\alpha^{\prime\prime}(z)\right)r_\perp^4, 
\label{Hoprime}  \\ 
\hat{\mathcal{H}}_o^{(\hbar)} 
   & \approx &   \frac{\hbar^2}{8p_0}\left({\alpha^\prime(z)}^2
              - 2\alpha(z)\alpha^{\prime\prime}(z)\right)r_\perp^2  
              + \frac{\hbar^2}{32p_0}\left(\alpha^{\prime\prime}(z)^2  
              - \alpha^\prime(z)\alpha^{\prime\prime\prime}(z)\right)r_\perp^4, 
\label{Hohbar}  \\ 
\hat{\mathcal{H}}_{o,\sigma}^{(\hbar)} 
   & \approx & \frac{3\hbar^2}{64p_0}\alpha^\prime(z)\alpha^{\prime\prime}(z) 
               r_\perp^2\left(y\Sigma_x - x\Sigma_y\right) + \ldots,     
\label{Hosigma}
\end{eqnarray} 
with  
\begin{equation} 
	\alpha(z) = \frac{qB(z)}{2p_0}, \quad 
	\vec{\Sigma} = \left(\begin{array}{cc}  
		\vec{\sigma} & \mathbf{0} \\ 
		\mathbf{0} & \vec{\sigma}                          
	\end{array}\right).     
\label{alpha}
\end{equation} 
Note that $\hat{\mathcal{H}}_o$ depends explicitly on $z$ through $\alpha(z)$.  It should be 
noted that the design momentum $p_0$ in $\hat{\mathcal{H}}_o$ can be nonrelativistic or 
relativistic.  In the relativistic case, with $\gamma = 1/\sqrt{1-v_0^2/c^2}$ and 
$E = \gamma mc^2$, $p_0 = \sqrt{E^2-m^2c^4}/c = \gamma mv_0$.  In the nonrelativistic case, 
with $\gamma \approx [1 + (v_0^2/2c^2)]$ and $E \approx mc^2 + (mv_0^2/2)$, $p_0 \approx mv_0$.  
Thus, there is no nonrelativistic approximation for $\hat{\mathcal{H}}_o$ except for choosing 
$p_0$ appropriately.  The approximations for $\hat{\mathcal{H}}_o$ depend only on whether the 
beam is paraxial or nonparaxial.  Equation (\ref{QBOevoln}) is the quantum electron beam 
optical evolution equation with $\hat{\mathcal{H}}_o$ as the quantum electron beam optical Hamiltonian.  

Now, we are interested in the $z$-evolution of $\underline{\psi}\left(\vec{r}_\perp,z\right)$ 
according to (\ref{QBOevoln}).  Note that the equations (\ref{probatt}) and (\ref{Psi}) imply 
that 
\begin{equation} 
\underline{\psi}\left(\vec{r}_\perp,z\right)^\dagger
\underline{\psi}\left(\vec{r}_\perp,z\right) 
   = \sum_{j=1}^{4}\psi_j^*\left(\vec{r}_\perp,z\right)\psi_j\left(\vec{r}_\perp,z\right),   
\end{equation} 
is the probability of finding the electron at the position $(x,y)$ in the transverse plane at 
the point $z$ on the optic axis at any time $t$ and 
$\int d^2r_\perp\,\underline{\psi}\left(\vec{r}_\perp,z\right)^\dagger
\underline{\psi}\left(\vec{r}_\perp,z\right)$
is the probability of finding the electron somewhere in the plane perpendicular to the axis at 
the point $z$.  Here we are writing $\int d^2r_\perp$ for $\int_{-\infty}^{\infty}\int_{-\infty}^{\infty}dxdy$.  The probability of finding the electron 
somewhere in space (\ref{probatt}) at any time $t$ is always one and is conserved since it 
cannot disappear from space.  But, the probability of finding the electron somewhere in the 
plane perpendicular to the axis at a point $z$ need not have the same value {\em i.e.}, be 
conserved, along the $z$-axis.  This explains why $\hat{\mathcal{H}}$ in (\ref{zevoln}) is  non-Hermitian.  However, if we consider the probability for a beam electron to be lost during 
the transit through the system to be negligible we can assume that the effect of the 
non-Hermitian terms in the Hamiltonian to be negligible and drop them.  This is what we have 
done in arriving at the quantum electron beam optical Hamiltonian $\hat{\mathcal{H}}_o$ in 
(\ref{DiracHo}) which is Hermitian.  With Hermitian $\hat{\mathcal{H}}_o$ the $z$-evolution of $\underline{\psi}\left(\vec{r}_\perp,z\right)$ is unitary so that we can normalize the wave 
function $\underline{\psi}\left(\vec{r}_\perp,z\right)$ as   
\begin{equation} 
\int d^2r_\perp\,\underline{\psi}\left(\vec{r}_\perp,z\right)^\dagger
          \underline{\psi}\left(\vec{r}_\perp,z\right) = 1,\ \ \mbox{at any}\ z.       	
\end{equation} 
This means that we are considering the probability of finding the electron somewhere in the 
transverse $(x,y)$ plane at a point $z$ on the axis is same for any $z$ and we have normalized 
this value to be one.  

In $\hat{\mathcal{H}}_o$ in (\ref{DiracHo}), $\hat{\mathcal{H}}_{o,p}$ in (\ref{Hop}) is 
responsible for the paraxial behavior, and $\hat{\mathcal{H}}^\prime_o$ in (\ref{Hoprime}) 
gives rise to third-order aberrations.  The part $\hat{\mathcal{H}}_o^{(\hbar)}$ in 
(\ref{Hohbar}) leads to explicitly $\hbar$-dependent modifications of the paraxial behavior 
and aberrations.  The part $\hat{\mathcal{H}}_{o,\sigma}^{(\hbar)}$ in (\ref{Hosigma}) is the 
matrix part originating from the spin of the electron, or the $4$-component nature of the 
electron wave function, besides being explicitly $\hbar$-dependent.  Since, the terms in both $\hat{\mathcal{H}}_o^{(\hbar)}$ and $\hat{\mathcal{H}}_{o,\sigma}^{(\hbar)}$ are proportional 
to $\hbar$ and powers of $\hbar$ their contributions are tiny and can be dropped from $\hat{\mathcal{H}}_o$.  If we drop these $\hbar$-dependent terms the remaining part of 
$\hat{\mathcal{H}}_o$, namely 
$\left(\hat{\mathcal{H}}_{o,p} + \hat{\mathcal{H}}^\prime_o\right)$, is the `classical' scalar 
term which acts on each component of $\underline{\psi}\left(\vec{r}_\perp,z\right)$ 
independently.  This implies that we can take, effectively, the electron wave function to be 
a scalar {\em i.e.}, a single-component wave function like the nonrelativistic Schr\"{o}dinger 
wave function, ignoring the electron spin or treating it as a spectator.  Thus, we shall take 
\begin{equation} 
i\hbar\frac{\partial\psi\left(\vec{r}_\perp,z\right)}{\partial z} 
   = \hat{\mathcal{H}}_{o,c}\psi\left(\vec{r}_\perp,z\right),  \quad 
\hat{\mathcal{H}}_{o,c} 
   = \hat{\mathcal{H}}_{o,p} + \hat{\mathcal{H}}^\prime_o,  
\label{scalarQBOevoln} 
\end{equation} 
as the quantum electron beam optical evolution equation with 
$\psi\left(\vec{r}_\perp,z\right)$ as the scalar wave function representing the beam electron 
and $\hat{\mathcal{H}}_{o,c}$ as the quantum electron beam optical Hamiltonian.  

One may wonder whether it was necessary at all to start with the Dirac equation to obtain 
the scalar wave function description and whether it would not have been enough to start with 
the scalar relativistic Klein-Gordon equation.  It is true that if we are to drop all the 
$\hbar$-dependent scalar and matrix terms from the quantum electron beam optical Hamiltonian 
there would be no difference whether we start with the Dirac equation or the Klein-Gordon 
equation.  However, if we are to include the $\hbar$-dependent scalar terms then $\hat{\mathcal{H}}_o^{(\hbar)}$ in (\ref{Hohbar}) is found to be different from the 
corresponding term obtained when one starts with the Klein-Gordon equation (compare 
(\ref{Hohbar}) with Eq.(5.213) in \cite{Jagannathan2019}).  Thus, the 
signature of electron spin is seen even in the $\hbar$-dependent scalar part besides, obviously, 
in the matrix, or spin, part of the quantum electron beam optical Hamiltonian 
(see \cite{Jagannathan1996, Jagannathan2019} for more details).  

Foldy-Wouthuysen-like transformations are useful also in series expansions of the Klein-Gordon 
equation and the Helmholtz equation 
(see {\em e.g.}, \cite{Jagannathan1996, Khan1997, Jagannathan2019, Bjorken1994, Greiner2000}; 
see also \cite{Fishman2004} and references therein). The formalism of quantum charged particle 
beam  optics \cite{Jagannathan2019} has had a decisive influence on the development of new 
formalisms of scalar light beam optics based on the Helmholtz equation and vector light beam 
optics, incorporating polarization, based on the Maxwell equations 
\cite{Khan2002b, Khan2006, Khan2008, Khan2016a, Khan2016b, Khan2016c, Khan2017a}.  The use of 
quantum methodologies in light beam optics has provided new insights on Hamilton's 
optico-mechanical analogy and also extends it into the wavelength-dependent regime 
\cite{Khan2017b}. 

\section{Equations of motion for the quantum averages of observables}
Let us look at the evolution of the observables of the beam electron defined in the plane 
perpendicular to the optical axis as it transits through the system.  We can take over the usual quantum mechanical formalism for time evolution by replacing time $t$ by $z$.  In general, for 
an observable, say  $O\left(\vec{r}_\perp,\vec{p}_\perp,z\right)$, represented by the 
corresponding Hermitian operator $\hat{O}\left(\vec{r}_\perp,\vec{\hat{p}}_\perp,z\right)$,   
\begin{equation} 
\langle\hat{O}\rangle(z) 
   = \int d^2r_\perp\,\psi^*\left(\vec{r}_\perp,z\right) 
                   \hat{O}\left(\vec{r}_\perp,\vec{\hat{p}}_\perp,z\right) 
                   \psi\left(\vec{r}_\perp,z\right)   
   = \left\langle\psi(z)\right|\hat{O}\left|\psi(z)\right\rangle 
\end{equation}
gives its expectation value, or average, in the transverse plane at $z$.  Since $\hat{O}$ is 
Hermitian $\langle\hat{O}\rangle(z)$ is real.  We can take $\langle\hat{O}\rangle(z)$ as 
corresponding to the value of the classical variable $O$ in the transverse plane at $z$, 
following the Ehrenfest theorem (see {\em e.g.}, \cite{Greiner2001, Griffiths2018}, and also \cite{Jagannathan1996, Jagannathan2019}).  In particular, we are interested in the $z$-evolution 
of $\langle\vec{r}_\perp\rangle(z)$ and $\langle\vec{\hat{p}}_\perp\rangle(z)/p_0$, representing 
the classical ray coordinates $\vec{r}_\perp(z)$ and $d\vec{r}_\perp(z)/dz$, respectively.  To 
this end, we proceed as follows, borrowing the well known techniques of the study of 
time-evolution of quantum systems.  

We can write, as the formal solution of (\ref{scalarQBOevoln}),   
\begin{equation}
\psi\left(\vec{r}_\perp,z\right) 
   = \hat{U}\left(z,z_i\right)\psi\left(\vec{r}_\perp,z_i\right), 
\label{psizi2psiz}
\end{equation}
where $z_i$ and $z$ are the initial and final points of observation in the optic axis and $\hat{U}\left(z,z_i\right)$ is the $z$-evolution operator.  Substituting this expression 
for $\psi\left(\vec{r}_\perp,z\right)$ in (\ref{scalarQBOevoln}) it is seen that 
\begin{equation}
i\hbar\frac{\partial\hat{U}\left(z,z_i\right)}{\partial z} 
   = \hat{\mathcal{H}}_{o,c}(z)\hat{U}\left(z,z_i\right),  \qquad  
\hat{U}\left(z_i,z_i\right) = \hat{I},   
\label{Ueqn}
\end{equation}
where $\hat{I}$ is the identity operator.  Taking the Hermitian adjoint on both sides of 
(\ref{Ueqn}) we have 
\begin{eqnarray}
i\hbar\frac{\partial\hat{U}^\dagger\left(z,z_i\right)}{\partial z} 
   = -\hat{U}^\dagger\left(z,z_i\right)\hat{\mathcal{H}}_{o,c}(z),  \qquad 
\hat{U}^\dagger\left(z_i,z_i\right) = \hat{I},      
\label{Udageqn}
\end{eqnarray} 
where we have used the fact that $\hat{\mathcal{H}}_{o,c}(z)$ is Hermitian.  Then, multiplying both 
sides of (\ref{Ueqn}) from left by $\hat{U}^\dagger$ and both sides of (\ref{Udageqn}) from 
right by $\hat{U}$ and adding the two equations we get $i\hbar\partial(\hat{U}^\dagger\hat{U})/{\partial z} = 0$, showing that $\hat{U}^\dagger\hat{U}$ 
is independent of $z$.  Since  $\hat{U}^\dagger(z_i,z_i)\hat{U}(z_i,z_i) = I$, it follows that $\hat{U}^\dagger\hat{U} = I$ {\em i.e.}, $\hat{U}(z,z_i)$ is unitary.  From the equation $\hat{U}(\hat{U}^\dagger\hat{U}) = (\hat{U}\hat{U}^\dagger)\hat{U} = \hat{U}$ we can conclude 
that $\hat{U}\hat{U}^\dagger = I$.  From (\ref{Ueqn}), we have 
\begin{equation}
\int_{z_i}^{z}dz\,\frac{\partial\hat{U}\left(z,z_i\right)}{\partial z} 
   = -\frac{i}{\hbar}\int_{z_i}^{z}dz_1\,\hat{\mathcal{H}}_{o,c}\left(z_1\right)
      \hat{U}\left(z_1,z_i\right), 
\end{equation}
or 
\begin{equation}
\left.\hat{U}\left(z,z_i\right)\right|_{z_i}^{z} 
   =  \hat{U}\left(z,z_i\right) - I    
   = -\frac{i}{\hbar}\int_{z_i}^{z}dz_1\,\hat{\mathcal{H}}_{o,c}\left(z_1\right)
          \hat{U}\left(z_1,z_i\right).  
\end{equation} 
This leads to the formal solution 
\begin{equation}
\hat{U}\left(z,z_i\right)  
   = I - \frac{i}{\hbar}\int_{z_i}^{z}dz_1\,\hat{\mathcal{H}}_{o,c}\left(z_1\right)
         \hat{U}\left(z_1,z_i\right). 
\end{equation} 
Iterating this formal solution we get 
\begin{eqnarray}
\hat{U}\left(z,z_i\right)  
   & = & I - \frac{i}{\hbar}\int_{z_i}^{z}dz_1\,\hat{\mathcal{H}}_{o,c}\left(z_1\right)  
             \nonumber \\  
   &   & + \left(-\frac{i}{\hbar}\right)^2\int_{z_i}^{z}dz_2\int_{z_i}^{z_2}dz_1
                 \hat{\mathcal{H}}_{o,c}\left(z_2\right)\hat{\mathcal{H}}_{o,c}\left(z_1\right) 
                 \nonumber \\ 
   &   & + \left(-\frac{i}{\hbar}\right)^3\int_{z_i}^{z}dz_3\int_{z_i}^{z_3}dz_2
                 \int_{z_i}^{z_2}dz_1\,\hat{\mathcal{H}}_{o,c}\left(z_3\right)
                 \hat{\mathcal{H}}_{o,c}\left(z_2\right)
                 \hat{\mathcal{H}}_{o,c}\left(z_1\right) \nonumber \\ 
   &   & + \ldots.        
\label{zexp}
\end{eqnarray}
Note that $\hat{U}\left(z,z_i\right)$ becomes 
$\exp{(-i(z-z_i)\hat{\mathcal{H}}_{o,c}/\hbar)}$ if $\hat{\mathcal{H}}_{o,c}$ is 
independent of $z$.  Let us write, in general, 
\begin{equation}
\hat{U}\left(z,z_i\right) 
   = \mathsf{P}\left[\exp{\left(-\frac{i}{\hbar}
   	                 \int_{z_i}^{z}dz\,\hat{\mathcal{H}}_{o,c}(z)\right)}\right], 
\end{equation} 
where $\mathsf{P}[\cdots]$, say, the $z$-ordered exponential, represents symbolically the 
infinite series in the right hand side of (\ref{zexp}).  An equivalent expression for 
$\hat{U}\left(z,z_i\right)$ is given by the Magnus formula \cite{Magnus1954} (see also 
\cite{Jagannathan1996, Jagannathan2019, Blanes2009}): 
\begin{eqnarray}
\hat{U}\left(z,z_i\right) 
   & = & \exp{\left(-\frac{i}{\hbar}\hat{T}\left(z,z_i\right)\right)},  \nonumber \\ 
\hat{T}\left(z,z_i\right)
   & = & \int_{z_i}^{z}dz_1\,\hat{\mathcal{H}}_{o,c}\left(z_1\right)  \nonumber \\ 
   &   & + \frac{1}{2}\left(-\frac{i}{\hbar}\right)\int_{z_i}^{z}dz_2\int_{z_i}^{z_2} 
                 dz_1\,\left[\hat{\mathcal{H}}_{o,c}\left(z_2\right),
                       \hat{\mathcal{H}}_{o,c}\left(z_1\right)\right]  \nonumber \\  
   &   & + \frac{1}{6}\left(-\frac{i}{\hbar}\right)^2 
                 \int_{z_i}^{z}dz_3\int_{z_i}^{z_3}dz_2\int_{z_i}^{z_2}dz_1 
           \left\{\left[\left[\hat{\mathcal{H}}_{o,c}
                 \left(z_3\right),\hat{\mathcal{H}}_{o,c}\left(z_2\right)\right],
                 \hat{\mathcal{H}}_{o,c}\left(z_1\right)\right] \right. \nonumber \\ 
   &   & \qquad\qquad\qquad\qquad\qquad\qquad\qquad \left. + \left[\left[\hat{\mathcal{H}}_{o,c}
                 \left(z_1\right),\hat{\mathcal{H}}_{o,c}\left(z_2\right)\right],
                 \hat{\mathcal{H}}_{o,c}\left(z_3\right)\right]\right\} \nonumber \\ 
   &   & + \ldots.  
\end{eqnarray} 
Note that $\hat{T}(z,z_i)$ is Hermitian when $\hat{\mathcal{H}}_{o,c}$ is Hermitian such that $\hat{U}(z,z_i)$ is unitary.  

Now, we can write the average value $\langle\hat{O}\rangle(z)$ for any $\hat{O}$ as 
\begin{eqnarray}
\langle\hat{O}\rangle(z) 
   & = & \int d^2r_\perp\,\left(\hat{U}\left(z,z_i\right)
                          \psi\left(\vec{r}_\perp,z_i\right)\right)^*
         \hat{O}\hat{U}\left(z,z_i\right)\psi\left(\vec{r}_\perp,z_i\right)  \nonumber \\ 
   & = & \int d^2r_\perp\,\psi\left(\vec{r}_\perp,z_i\right)^*
         \hat{U}^\dagger\left(z,z_i\right)\hat{O}\hat{U}\left(z,z_i\right)
         \psi\left(\vec{r}_\perp,z_i\right)  \nonumber \\ 
   & = & \left\langle\psi\left(z_i\right)\right|\hat{U}^\dagger\left(z,z_i\right)\hat{O}
         \hat{U}\left(z,z_i\right)\left|\psi\left(z_i\right)\right\rangle.
\label{Oz}
\end{eqnarray} 
For any observable the equation of motion along the $z$-axis is given by 
\begin{eqnarray}
\frac{d\langle\hat{O}\rangle(z)}{dz} 
   & = & \left\langle\psi\left(z_i\right)\left|
         \frac{\partial\hat{U}^\dagger\left(z,z_i\right)}{\partial z}\hat{O}
         \hat{U}\left(z,z_i\right)\right|\psi\left(z_i\right)\right\rangle  \nonumber \\ 
   &   & + \left\langle\psi\left(z_i\right)\left|
                 \hat{U}^\dagger\left(z,z_i\right)\hat{O}
                 \frac{\partial\hat{U}\left(z,z_i\right)}
                {\partial z}\right|\psi\left(z_i\right)\right\rangle  \nonumber \\ 
   &   & + \left\langle\psi\left(z_i\right)\left|
                 \hat{U}^\dagger\left(z,z_i\right)\frac{\partial\hat{O}}{\partial z}
                 \hat{U}\left(z,z_i\right)\right|\psi\left(z_i\right)\right\rangle  
                 \nonumber \\ 
   & = &  \frac{i}{\hbar}\left\langle\psi\left(z_i\right)\left|\hat{U}^\dagger
          \left(z,z_i\right)\left[\hat{\mathcal{H}}_{o,c},\hat{O}\right] 
          \hat{U}\left(z,z_i\right)\right|
          \psi\left(z_i\right)\right\rangle  \nonumber \\ 
   &   & + \left\langle\psi\left(z_i\right)\left|
                 \hat{U}^\dagger\left(z,z_i\right)\frac{\partial\hat{O}}{\partial z}
                 \hat{U}\left(z,z_i\right)\right|\psi\left(z_i\right)\right\rangle  
                 \nonumber \\ 
   & = &   \frac{i}{\hbar}\left\langle\left[\hat{\mathcal{H}}_{o,c},\hat{O}\right]
           \right\rangle(z) + \left\langle\frac{\partial\hat{O}}{\partial z}
           \right\rangle(z), 
\label{HEq} 
\end{eqnarray} 
where we have used (\ref{Ueqn}), (\ref{Udageqn}) and (\ref{Oz}).  

The equation (\ref{scalarQBOevoln}) is the quantum electron beam optical equation of motion 
for the wave function $\psi\left(\vec{r}_\perp,z\right)$ in the Schr\"{o}dinger picture in 
which the wave function changes with $z$ and an observable of the electron does not change 
with $z$ in the absence of any explicit $z$-dependence.  In the Heisenberg picture the wave 
function remains fixed, $\psi\left(\vec{r}_\perp,z_i\right)$, and an observable of the 
electron changes with $z$ as 
\begin{equation}
\hat{O}(z) = \hat{U}^\dagger\left(z,z_i\right)\hat{O}\hat{U}\left(z,z_i\right),   
\end{equation} 
so that the average value of the observable is given by 
\begin{equation}
\langle\hat{O}\rangle(z) 
   = \left\langle\psi\left(z_i\right)\left|\hat{O}(z)\right|\psi\left(z_i\right)\right\rangle.     
\end{equation}
For the Heisenberg observable $\hat{O}(z)$ we get the equation of motion from (\ref{HEq}) as 
\begin{equation}
\frac{d\hat{O}(z)}{dz} 
   =    \frac{i}{\hbar}\left[\hat{\mathcal{H}}_{o,c}(z),\hat{O}(z)\right]
         + \left(\frac{\partial\hat{O}}{\partial z}\right)(z)  
   =    \frac{1}{i\hbar}\left[\hat{O}(z),\hat{\mathcal{H}}_{o,c}(z)\right]
         + \left(\frac{\partial\hat{O}}{\partial z}\right)(z).    
\label{HEqM}
\end{equation}
This is seen to correspond to the Hamilton equation of motion for the classical observables 
\begin{equation}
\frac{dO}{dz} = \left\{O,\mathcal{H}_{o,c}\right\}+\frac{\partial O}{\partial z}, 
\label{dOdzclassical}
\end{equation}    
where $\{\,,\,\}$ is the classical Poisson bracket, $\mathcal{H}_{o,c}$ is the classical 
electron beam optical Hamiltonian which is obtained by replacing in $\hat{\mathcal{H}}_{o,c}$ 
the quantum operators $\vec{r}_\perp$ and $\vec{\hat{p}}_\perp$ by the classical variables 
$\vec{r}_\perp$ and $\vec{p}_\perp$, respectively, and the quantum operator $\hat{O}$ is 
replaced by the classical observable $O$.  Dirac's `quantum $\longleftrightarrow$ classical' correspondence rule 
\begin{equation} 
\frac{1}{i\hbar}\left[\hat{A},\hat{B}\right] \longleftrightarrow \{A,B\}, 
\label{QCcorrespondence}
\end{equation} 
takes (\ref{HEqM}) to (\ref{dOdzclassical}).  As suggested by this, we can obtain the 
`classical part' of the quantum electron beam optical Hamiltonian $\hat{\mathcal{H}}_o$ 
{\em i.e.}, $\hat{\mathcal{H}}_{o,c} = \hat{\mathcal{H}}_{o,p} + \hat{\mathcal{H}}_o^\prime$, 
without the terms dependent on $\hbar$ and spin matrices, starting with the classical electron 
beam optical Hamiltonian and replacing the classical variables in it by the corresponding 
Hermitian quantum operators and proper  symmetrization to make the resulting Hamiltonian 
operator Hermitian.  

When the observable is not explicitly dependent on $z$ the equation of motion (\ref{HEq}) 
becomes 
\begin{equation} 
\frac{d\langle\hat{O}\rangle(z)}{dz} 
   = \frac{i}{\hbar}\left\langle\left[\hat{\mathcal{H}}_{o,c}\,,\,\hat{O}\right]\right\rangle(z).  
\end{equation}
We are particularly interested in the equations of motion for $x$, $y$, $\hat{p}_x/p_0$ and $\hat{p}_y/p_0$ which are independent of $z$.  Let us write 
$\langle x\rangle(z) = \langle x\rangle$, 
$\langle\hat{p}_x\rangle(z) = \langle\hat{p}_x\rangle$, etc., for the sake of simplicity of 
notation.  The desired equations of motion are as follows:      
\begin{eqnarray}
\frac{d\left\langle x\right\rangle}{dz}   
   & = &   \frac{1}{p_0}\langle\hat{p}_x\rangle + \alpha(z)\langle y\rangle 
         + \frac{1}{2p_0^3}\left\langle\hat{p}_\perp^2\hat{p}_x\right\rangle 
         + \frac{\alpha(z)}{2p_0^2}\left(\left\langle\hat{p}_\perp^2y\right\rangle  
         - 2\left\langle\hat{p}_x\hat{L}_z\right\rangle\right)  \nonumber \\ 
   &   & - \frac{\alpha^2(z)}{4p_0}\left(\left\langle x\vec{r}_\perp\cdot
           \vec{\hat{p}}_\perp\right\rangle\left\langle\vec{r}_\perp\cdot
           \vec{\hat{p}}_\perp x\right\rangle 
         + \left\langle x\vec{\hat{p}}_\perp\cdot\vec{r}_\perp
           \right\rangle + \left\langle\vec{\hat{p}}_\perp\cdot\vec{r}_\perp x
           \right\rangle\right) \nonumber \\ 
   &   & + \frac{3\alpha^2(z)}{4p_0}\left(\left\langle r_\perp^2p_x\right\rangle 
         + \left\langle p_xr_\perp^2\right\rangle\right) 
         - \frac{1}{8}\left(\alpha^{\prime\prime}(z)-4\alpha^3(z)\right)
           \left\langle yr_\perp^2\right\rangle,  
\label{EqMx} \\  
\frac{d\left\langle y\right\rangle}{dz} 
   & = &   \frac{1}{p_0}\langle\hat{p}_y\rangle-\alpha(z)\langle x\rangle  
         + \frac{1}{2p_0^3}\left\langle\hat{p}_\perp^2\hat{p}_y\right\rangle   
         - \frac{\alpha(z)}{2p_0^2}\left(\left\langle\hat{p}_\perp^2x\right\rangle    
         + 2\left\langle\hat{p}_y\hat{L}_z\right\rangle\right)  \nonumber \\ 
   &   & - \frac{\alpha^2(z)}{4p_0}\left(\left\langle y\vec{r}_\perp\cdot 
           \vec{\hat{p}}_\perp\right\rangle + \left\langle\vec{r}_\perp\cdot
           \vec{\hat{p}}_\perp y\right\rangle 
         + \left\langle y\vec{\hat{p}}_\perp\cdot\vec{r}_\perp
           \right\rangle + \left\langle\vec{\hat{p}}_\perp\cdot\vec{r}_\perp y  
           \right\rangle\right)  \nonumber \\ 
   &   & + \frac{3\alpha^2(z)}{4p_0}\left(\left\langle r_\perp^2p_y\right\rangle
         + \left\langle p_yr_\perp^2\right\rangle\right)  
         + \frac{1}{8}\left(\alpha^{\prime\prime}(z)-4\alpha^3(z)\right)
           \left\langle xr_\perp^2\right\rangle,                  
\label{EqMy}  \\  
\frac{1}{p_0}\frac{d\langle\hat{p}_x\rangle}{dz}  
   & = &  -\alpha^2(z)\left\langle x\right\rangle
         + \frac{\alpha(z)}{p_0}\langle\hat{p}_y\rangle 
         + \frac{\alpha(z)}{2p_0^3}\left\langle\hat{p}_\perp^2\hat{p}_y\right\rangle  
           \nonumber \\  
   &   & + \frac{\alpha^2(z)}{4p_0^2}\left(\left\langle\hat{p}_x\vec{r}_\perp\cdot
           \vec{\hat{p}}_\perp\right\rangle+\left\langle\vec{r}_\perp\cdot
           \vec{\hat{p}}_\perp\hat{p}_x\right\rangle 
         + \left\langle\hat{p}_x\vec{\hat{p}}_\perp\cdot\vec{r}_\perp\right\rangle 
         + \left\langle\vec{\hat{p}}_\perp\cdot\vec{r}_\perp\hat{p}_x\right\rangle\right)  
           \nonumber \\ 
   &   & - \frac{3\alpha^2(z)}{4p_0^2}\left(\left\langle x\hat{p}_\perp^2\right\rangle
         + \left\langle\hat{p}_\perp^2x\right\rangle\right)  \nonumber \\  
   &   & - \frac{1}{8p_0}\left(\left(\alpha^{\prime\prime}(z) - 4\alpha^3(z)\right) 
           \left(2\left\langle\hat{L}_zx\right\rangle
         + \left\langle\hat{p}_yr_\perp^2\right\rangle\right)\right)  \nonumber \\    
   &   & - \frac{1}{2}\left(\alpha^4(z) - \alpha(z)\alpha^{\prime\prime}(z)\right)
           \left\langle r_\perp^2x\right\rangle, 
\label{EqMpx}  \\ 
\frac{1}{p_0}\frac{d\langle\hat{p}_y\rangle}{dz} 
   & = &  -\alpha^2(z)\left\langle y\right\rangle
          - \frac{\alpha(z)}{p_0}\langle\hat{p}_x\rangle  
          - \frac{\alpha(z)}{2p_0^3}\left\langle\hat{p}_\perp^2\hat{p}_x\right\rangle  
            \nonumber \\ 
   &   & + \frac{\alpha^2(z)}{4p_0^2}\left(\left\langle\hat{p}_y\vec{r}_\perp\cdot
           \vec{\hat{p}}_\perp\right\rangle+\left\langle\vec{r}_\perp\cdot
           \vec{\hat{p}}_\perp\hat{p}_y\right\rangle 
         + \left\langle\hat{p}_y\vec{\hat{p}}_\perp\cdot\vec{r}_\perp\right\rangle 
         + \left\langle\vec{\hat{p}}_\perp\cdot\vec{r}_\perp\hat{p}_y\right\rangle\right)  
           \nonumber \\ 
   &   & - \frac{3\alpha^2(z)}{4p_0^2}\left(\left\langle y\hat{p}_\perp^2\right\rangle
         + \left\langle\hat{p}_\perp^2y\right\rangle\right)  \nonumber \\ 
   &   & - \frac{1}{8p_0}\left(\left(\alpha^{\prime\prime}(z) - 4\alpha^3(z)\right) 
           \left(2\left\langle\hat{L}_zy\right\rangle
         - \left\langle\hat{p}_xr_\perp^2\right\rangle\right)\right)  \nonumber \\ 
   &   & - \frac{1}{2}\left(\alpha^4(z)-\alpha(z)\alpha^{\prime\prime}(z)\right)
           \left\langle r_\perp^2y\right\rangle.   
\label{EqMpy}
\end{eqnarray} 
Note that all the terms on the right hand sides of the above equations are real, as should 
be.  A term like $(\langle\hat{p}_\perp^2y\rangle-2\langle\hat{p}_x\hat{L}_z\rangle)$ 
may not look explicitly real, but can be seen to be real by checking that the operator 
$(\hat{p}_\perp^2y-2\hat{p}_x\hat{L}_z)$ is Hermitian.  

The following should be noted.  Even after we have dropped the explicitly $\hbar$-dependent 
terms from the Hamiltonian $\hat{\mathcal{H}}_o$ the effect of quantum mechanics on the 
equations of motion can be seen in the terms not linear in $\langle x\rangle$, 
$\langle y\rangle$, $\langle\hat{p}_x\rangle$, and 
$\langle\hat{p}_y\rangle$.  For example, let us consider the term 
$\left\langle x^3\right\rangle$ (part of the term 
$\left\langle r_\perp^2x\right\rangle 
= \left\langle x^3\right\rangle+\left\langle y^2x\right\rangle$) and express it in terms of 
$\langle x\rangle$ which would correspond to the result of a position measurement.  Let  
$\delta x = x-\langle x\rangle$.  Then, $\langle\delta x\rangle = 0$.  We can write 
\begin{eqnarray}  
\left\langle x^3\right\rangle 
   & = &   \left\langle\left(\langle x\rangle + \delta x\right)^3\right\rangle    
     =     \left\langle\left(\langle x\rangle^3 + 3\langle x\rangle^2\delta x 
         + 3\langle x\rangle(\delta x)^2 + (\delta x)^3\right)\right\rangle \nonumber \\ 
   & = &   \langle x\rangle^3 + 3\langle x\rangle\left\langle(\delta x)^2\right\rangle 
         + \left\langle(\delta x)^3\right\rangle.  
\end{eqnarray} 
This shows that the quantum electron beam optical equations of motion differ from the 
corresponding classical electron beam optical equations of motion since a term like 
$\left\langle x^3\right\rangle$ cannot be replaced by $\langle x\rangle^3$.  The difference 
is seen to depend on terms like $(\Delta x)^2 = \left\langle(\delta x)^2\right\rangle$, 
the square of the uncertainty in $x$, and higher order central moments like 
$\left\langle(\delta x)^3\right\rangle$.  Thus, in principle, the quantum uncertainties will 
influence the equations of motion for the position and momentum of the beam electron and 
hence influence the performance of the system.  For example, when an aperture is introduced 
in the path of the beam to limit the transverse momentum spread one will be introducing 
uncertainties in the position coordinates, $\Delta x$ and $\Delta y$, and hence the 
corresponding momentum uncertainties, $\Delta p_x \geq \hbar/2\Delta x$ and 
$\Delta p_y \geq \hbar/2\Delta y$, in accordance with the Heisenberg uncertainty principle.  
This is uncontrollable.  However, in practice, these quantum effects might be too tiny to be 
of any significance in the present-day electron optical systems.  This explains the remarkable 
success of classical electron optics.    

\section{Paraxial approximation}
In the equations of motion (\ref{EqMx}-\ref{EqMpy}) the terms on the right hand side linear in 
$\langle x\rangle$, $\langle y\rangle$, $\langle\hat{p}_x\rangle$, 
and $\langle\hat{p}_y\rangle$, are the result of $\hat{\mathcal{H}}_{o,p}$, the paraxial 
part of $\hat{\mathcal{H}}_{o,c}$, which is quadratic in $\vec{r}_\perp$ and $\vec{\hat{p}}_\perp$.  
The terms not linear in $\langle x\rangle$, $\langle y\rangle$, $\langle\hat{p}_x\rangle$, and $\langle\hat{p}_y\rangle$, are the result of $\hat{\mathcal{H}}_o^\prime$.  Now, let us keep 
only the terms linear in $\langle x\rangle$, $\langle y\rangle$, $\langle\hat{p}_x\rangle$, and $\langle\hat{p}_y\rangle$, on the right hand side of the equations of motion. In other words, 
let us take 
\begin{equation} 
i\hbar\frac{\partial\psi\left(\vec{r}_\perp,z\right)}{\partial z} 
   = \hat{\mathcal{H}}_{o,p}\psi\left(\vec{r}_\perp,z\right),     
\label{paraxevolneq}
\end{equation}  
as the quantum paraxial electron beam optical evolution equation, with the quantum paraxial electron beam optical Hamiltonian $\hat{\mathcal{H}}_{o,p}$ given by (\ref{Hop}).  This paraxial, or linear, approximation leads to the paraxial equations of motion,   
\begin{equation}
\frac{d}{dz}\left(\begin{array}{c}
                  \langle x\rangle(z) \\ 
                  \langle y\rangle(z) \\ 
                  \frac{1}{p_0}\langle\hat{p}_x\rangle(z) \\ 
                  \frac{1}{p_0}\langle\hat{p}_y\rangle(z) 
                  \end{array}\right)  
   = \left(\begin{array}{cccc} 
           0 & \alpha(z) & 1 & 0 \\
           -\alpha(z) & 0 & 0 & 1 \\ 
           -\alpha^2(z) & 0 & 0 & \alpha(z) \\ 
            0 & -\alpha^2(z) & -\alpha(z) & 0 
            \end{array}\right)  
      \left(\begin{array}{c}
            \langle x\rangle(z) \\ 
            \langle y\rangle(z) \\ 
            \frac{1}{p_0}\langle\hat{p}_x\rangle(z) \\ 
            \frac{1}{p_0}\langle\hat{p}_y\rangle(z) 
            \end{array}\right),      
\label{paraxEq}
\end{equation} 
following from (\ref{EqMx}-\ref{EqMpy}).  Let us write this equation as 
\begin{eqnarray}
\frac{d}{dz}\left(\begin{array}{c}
                  \left\langle\vec{r}_\perp\right\rangle(z) \\ 
                  \frac{1}{p_0}\left\langle\vec{\hat{p}}_\perp\right\rangle(z)  
                  \end{array}\right)  
   & = & \left(\begin{array}{cc}
               \rho(z) & \mathbf{1} \\ 
              -\alpha^2(z)\mathbf{1} & \rho(z) 
               \end{array}\right)
         \left(\begin{array}{c}
               \left\langle\vec{r}_\perp\right\rangle(z) \\ 
               \frac{1}{p_0}\left\langle\vec{\hat{p}}_\perp\right\rangle(z)  
               \end{array}\right),  \nonumber \\ 
   &   & \qquad\qquad \mbox{with}\ \ \rho(z) = \left(\begin{array}{cc} 
                                                    0 & \alpha(z) \\ 
                                                   -\alpha(z) & 0  
                                                    \end{array}\right).    
\label{ddzrp}
\end{eqnarray} 
Let us now introduce the Larmor $XYz$-coordinate frame with its $X$ and $Y$ axes rotating along 
the $z$-axis, coinciding with the $z$-axis of the laboratory frame, at the $z$-dependent rate $d\theta(z)/dz = \alpha(z)$ such that we can write, with 
$\theta\left(z,z_i\right) = \int_{z_i}^{z}dz\,\alpha(z)$,  
\begin{equation}
\left(\begin{array}{c}
      x \\ y 
      \end{array}\right) 
   = \left(\begin{array}{rr}
           \cos\theta\left(z,z_i\right) & \sin\theta\left(z,z_i\right) \\ 
          -\sin\theta\left(z,z_i\right) & \cos\theta\left(z,z_i\right)  
           \end{array}\right)
     \left(\begin{array}{c}
           X \\ Y  
           \end{array}\right) 
   = \mathsf{R}\left(z,z_i\right)
     \left(\begin{array}{c} 
           X \\ Y  
           \end{array}\right).     
\label{xy2XY}
\end{equation} 
Then, 
\begin{equation}
\left(\begin{array}{c}
      \hat{p}_x \\ \hat{p}_y 
      \end{array}\right)
   = \left(\begin{array}{c}
         -i\hbar\partial/\partial x \\ -i\hbar\partial/\partial y  
           \end{array}\right)   
   = \mathsf{R}\left(z,z_i\right)
     \left(\begin{array}{c}
         -i\hbar\partial/\partial X \\ -i\hbar\partial/\partial Y   
           \end{array}\right)   
   = \mathsf{R}\left(z,z_i\right)
     \left(\begin{array}{c}
           \hat{P}_{X} \\ \hat{P}_{Y} 
           \end{array}\right).  
\label{pxpy2PXPY}
\end{equation} 
Let us recall that $z_i$ and $z$ are the initial and final points of observation in the optic 
axis.   Note that $\alpha(z) = qB(z)/2p_0 = qB(z)/2\gamma mv_0 = \omega_L(z)/v_0$ where 
$\omega_L(z)$ is the instantaneous Larmor frequency at $z$.  Now, we have 
\begin{equation}
\left(\begin{array}{c}
      \left\langle\vec{r}_\perp\right\rangle(z) \\  
      \frac{1}{p_0}\left\langle\vec{\hat{p}}_\perp\right\rangle(z)  
      \end{array}\right)  
   = \left(\begin{array}{cc}
           \mathsf{R}\left(z,z_i\right) & \mathbf{0} \\
           \mathbf{0} & \mathsf{R}\left(z,z_i\right)
           \end{array}\right) 
     \left(\begin{array}{c}
           \left\langle\vec{R}_\perp\right\rangle(z) \\ 
           \frac{1}{p_0}\left\langle\vec{\hat{P}}_\perp\right\rangle(z)  
           \end{array}\right).  
\label{rp2RP} 
\end{equation} 
This implies that, with $dR\left(z,z_i\right)/dz = \rho(z)R\left(z,z_i\right)$,  
\begin{eqnarray}
\frac{d}{dz}\left(\begin{array}{c}
                           \left\langle\vec{r}_\perp\right\rangle(z) \\ 
                           \frac{1}{p_0}\left\langle\vec{\hat{p}}_\perp\right\rangle(z)  
                           \end{array}\right)  
   & = & \left(\begin{array}{cc}
               \rho(z)\mathsf{R}\left(z,z_i\right) & \mathbf{0} \\ 
               \mathbf{0} & \rho(z)\mathsf{R}\left(z,z_i\right)
               \end{array}\right)
         \left(\begin{array}{c}
               \left\langle\vec{R}_\perp\right\rangle(z) \\ 
               \frac{1}{p_0}\left\langle\vec{\hat{P}}_\perp\right\rangle(z)  
               \end{array}\right)  \nonumber \\ 
   &   & \quad + \left(\begin{array}{cc}
                       \mathsf{R}\left(z,z_i\right) & \mathbf{0} \\
                       \mathbf{0} & \mathsf{R}\left(z,z_i\right)
                       \end{array}\right)\frac{d}{dz}
                 \left(\begin{array}{c}
                       \left\langle\vec{R}_\perp\right\rangle(z) \\ 
                       \frac{1}{p_0}\left\langle\vec{\hat{P}}_\perp\right\rangle(z)  
                       \end{array}\right).  \nonumber \\  
   &   & 
\label{ddzrp1}
\end{eqnarray}
From (\ref{ddzrp}) and (\ref{rp2RP}) we have 
\begin{eqnarray} 
\frac{d}{dz}\left(\begin{array}{c}
                  \left\langle\vec{r}_\perp\right\rangle(z) \\ 
                  \frac{1}{p_0}\left\langle\vec{\hat{p}}_\perp\right\rangle(z)  
                  \end{array}\right) 
   & = & \left(\begin{array}{cc}
               \rho(z) & \mathbf{1} \\ 
              -\alpha^2(z)\mathbf{1} & \rho(z) 
               \end{array}\right)
         \left(\begin{array}{cc}
               \mathsf{R}\left(z,z_i\right) & \mathbf{0} \\
               \mathbf{0} & \mathsf{R}\left(z,z_i\right)
               \end{array}\right)  \nonumber \\ 
   &   & \quad \times\left(\begin{array}{c}
                           \left\langle\vec{R}_\perp\right\rangle(z) \\ 
                           \frac{1}{p_0}\left\langle\vec{\hat{P}}_\perp\right\rangle(z)  
                           \end{array}\right).  
\label{ddzrp2}
\end{eqnarray}
Now, equating the right hand sides of (\ref{ddzrp1}) and (\ref{ddzrp2}) we get 
\begin{eqnarray}
\frac{d}{dz}\left(\begin{array}{c}
                  \left\langle\vec{R}_\perp\right\rangle(z) \\ 
                  \frac{1}{p_0}\left\langle\vec{\hat{P}}_\perp\right\rangle(z)  
                  \end{array}\right)  
   & = & \left(\begin{array}{cc}
               \mathsf{R}\left(z,z_i\right)^{-1} & \mathbf{0} \\
               \mathbf{0} & \mathsf{R}\left(z,z_i\right)^{-1}
               \end{array}\right)
         \left(\begin{array}{cc}
               \mathbf{0} & \mathbf{1} \\ 
              -\alpha^2(z)\mathbf{1} & \mathbf{0}  
               \end{array}\right)  \nonumber \\ 
   &   & \quad \times\left(\begin{array}{cc}
                            \mathsf{R}\left(z,z_i\right) & \mathbf{0} \\
                            \mathbf{0} & \mathsf{R}\left(z,z_i\right)
                            \end{array}\right)
                      \left(\begin{array}{c}
                            \left\langle\vec{R}_\perp\right\rangle(z) \\ 
                            \frac{1}{p_0}\left\langle\vec{\hat{P}}_\perp\right\rangle(z)  
                            \end{array}\right)  \nonumber \\ 
   & = & \left(\begin{array}{cc}
               0 & 1 \\ -\alpha^2(z) & 0  
               \end{array}\right)
         \left(\begin{array}{c}
               \left\langle\vec{R}_\perp\right\rangle(z) \\ 
               \frac{1}{p_0}\left\langle\vec{\hat{P}}_\perp\right\rangle(z)  
               \end{array}\right).     
\label{rptransfereqn}
\end{eqnarray} 
Thus, in the rotating $XYz$-coordinate system,  
\begin{equation} 
\frac{d\left\langle\vec{R}_\perp\right\rangle(z)}{dz} 
   = \frac{1}{p_0}\left\langle\vec{\hat{P}}_\perp\right\rangle(z) 
\label{Xeq}
\end{equation}
and 
\begin{equation}  
\frac{1}{p_0}\frac{d\left\langle\vec{\hat{P}}_\perp\right\rangle(z)}{dz}  
   = -\alpha^2(z)\left\langle\vec{R}_\perp\right\rangle(z),   
\label{Peq}
\end{equation} 
leading to 
\begin{equation}
\frac{d^2\left\langle\vec{R}_\perp\right\rangle(z)}{dz^2} 
   + \alpha^2(z)\left\langle\vec{R}_\perp\right\rangle(z) = 0, 
\label{paraxialEq} 
\end{equation} 
the well known classical paraxial equations of motion for the position coordinates of the 
electron in the rotating coordinate system.  Note that in the nonrelativistic case we can take 
$p_0^2 = 2meU$ where $U$ is the electric potential which has accelerated the electron.  Then 
we have $\alpha^2(z) = eB^2(z)/8mU$.  

The transfer map for the averages of $\left\langle\vec{R}_\perp\right\rangle$ and $\left\langle\vec{\hat{P}}_\perp\right\rangle$ between the transverse planes at $z_i$ and $z$ 
can be obtained by integrating (\ref{rptransfereqn}).  We can write the result as 
\begin{eqnarray}
\left(\begin{array}{c}
      \langle\vec{R}_\perp\rangle(z) \\ 
      \frac{1}{p_0}\left\langle\vec{\hat{P}}_\perp\right\rangle(z)  
      \end{array}\right)  
   & = & \left(\begin{array}{cc}
              g\left(z,z_i\right) & h\left(z,z_i\right) \\ 
             g^\prime\left(z,z_i\right) & h^\prime\left(z,z_i\right)
               \end{array}\right)
               \left(\begin{array}{c}
                     \langle\vec{R}_\perp\rangle\left(z_i\right) \\ 
                     \frac{1}{p_0}\left\langle\vec{\hat{P}}_\perp\right\rangle
                     \left(z_i\right)  
                     \end{array}\right)  \nonumber \\ 
   & = & \mathsf{M}\left(z,z_i\right)
                   \left(\begin{array}{c}
                         \langle\vec{r}_\perp\rangle\left(z_i\right) \\ 
                         \frac{1}{p_0}\left\langle\vec{\hat{r}}_\perp\right\rangle
                         \left(z_i\right)  
                         \end{array}\right),       
\label{Mzzi} 
\end{eqnarray} 
observing that the Larmor $XYz$ frame and the laboratory $xyz$ frame coincide at $z = z_i$.  
Note that the elements of the second row of $\mathsf{M}\left(z,z_i\right)$ are the 
$z$-derivatives of the elements of the first row because of (\ref{Xeq}).  We can recognize $g\left(z,z_i\right)$ and $h\left(z,z_i\right)$ as the two linearly independent solutions of 
the paraxial equation (\ref{paraxialEq}) corresponding to the initial conditions 
$\langle x\rangle\left(z_i\right) = 1$,
$\langle\hat{p}_x\rangle\left(z_i\right)/p_0 = d\langle x\rangle(z)/dz|_{z=z_i} = 0$ and 
$\langle x\rangle\left(z_i\right) = 0$,
$\langle\hat{p}_x\rangle\left(z_i\right)/p_0 = d\langle x\rangle(z)/dz|_{z=z_i} = 1$, 
respectively, and the same conditions for $y$ also.  Then, the two solutions have to satisfy 
the initial conditions 
\begin{equation}
g\left(z_i,z_i\right) = 1, \quad g^\prime\left(z_i,z_i\right) = 0,  \qquad 
h\left(z_i,z_i\right) = 0, \quad h^\prime\left(z_i,z_i\right) = 1.  
\label{ghinicondns}
\end{equation}
From (\ref{rp2RP}) and (\ref{Mzzi}) we get  
\begin{eqnarray}  
\left(\begin{array}{c} 
      \langle\vec{r}_\perp\rangle(z) \\  
      \frac{1}{p_0}\left\langle\vec{\hat{p}}_\perp\right\rangle(z)   
      \end{array}\right)  
   & = & \left(\begin{array}{cc}
              g\left(z,z_i\right)\mathsf{R}\left(z,z_i\right) & 
              h\left(z,z_i\right)\mathsf{R}\left(z,z_i\right) \\ 
             g^\prime\left(z,z_i\right)\mathsf{R}\left(z,z_i\right) & 
             h^\prime\left(z,z_i\right)\mathsf{R}\left(z,z_i\right)   
               \end{array}\right)  
         \left(\begin{array}{c}
               \langle\vec{r}_\perp\rangle\left(z_i\right) \\  
               \frac{1}{p_0}\left\langle\vec{\hat{p}}_\perp
               \right\rangle\left(z_i\right) 
               \end{array}\right).   
\label{paraxtrnsfrmap}
\end{eqnarray}
Thus, in terms of the two fundamental solutions $g\left(z,z_i\right)$ and 
$h\left(z,z_i\right)$ we can write 
\begin{equation} 
\langle\vec{r}_\perp\rangle(z) 
   = g\left(z,z_i\right)\mathsf{R}\left(z,z_i\right)
      \langle\vec{r}_\perp\rangle\left(z_i\right)  
     + \frac{1}{p_0}h\left(z,z_i\right)\mathsf{R}\left(z,z_i\right) 
       \left\langle\vec{\hat{p}}_\perp\right\rangle\left(z_i\right), 
\label{Rpath}
\end{equation}
to represent the `classical' trajectory of the quantum average of position of the electron 
entering the system with $\langle\vec{r}_\perp\rangle\left(z_i\right)$ and  $\left\langle\vec{\hat{p}}_\perp\right\rangle\left(z_i\right)$ as the average position and 
momentum in the transverse plane of entry at $z_i$. 

There are two ways of finding the fundamental solutions.  One is to solve the paraxial equation (\ref{paraxialEq}), a second order linear ordinary differential equation, analytically.  We will 
get two linearly independent solutions for the differential equation which can be combined 
linearly to get the two fundamental solutions with the required initial conditions.  An 
alternative method is as follows.  From (\ref{scalarQBOevoln}), (\ref{psizi2psiz}), (\ref{zexp}), 
and (\ref{rptransfereqn}), we have 
\begin{eqnarray}
\mathsf{M}\left(z,z_i\right)   
   & = & \mathsf{P}\left[\exp{\left(\int_{z_i}^{z}dz\,\mu(z)\right)}\right]  \nonumber \\ 
   & = & I + \int_{z_i}^{z}dz_1\,\mu\left(z_1\right)  
           + \int_{z_i}^{z}dz_2\int_{z_i}^{z_2}dz_1\,\mu\left(z_2\right)\mu\left(z_1\right) 
             \nonumber \\ 
   &   & + \int_{z_i}^{z}dz_3\int_{z_i}^{z_3}dz_2\int_{z_i}^{z_2}dz_1\,\mu\left(z_3\right)
               \mu\left(z_2\right)\mu\left(z_1\right)  \nonumber \\ 
   &   & + \ldots, \nonumber \\ 
   &   & \qquad \mbox{with}\ \ \mu(z) = \left(\begin{array}{cc} 
                                              0 & 1 \\ -\alpha^2(z) & 0   
                                              \end{array}\right).                      
\label{matrizant}
\end{eqnarray}
In the theory of linear ordinary differential equations this method of solving a system of 
linear differential equations is known as the Peano-Baker method and the above series expression 
for the transfer matrix $\mathsf{M}\left(z,z_i\right)$ is known as the Peano-Baker series 
(see {\em e.g.}, \cite{Pipes2014, Baake2011}).  Explicitly calculating the elements of the first 
row of $\mathsf{M}\left(z,z_i\right)$ we get 
\begin{eqnarray}
g\left(z,z_i\right)
   & = & 1 - \int_{z_i}^{z}dz_2\int_{z_i}^{z_2}dz_1\,\alpha^2\left(z_1\right)  \nonumber \\ 
   &   & + \int_{z_i}^{z}dz_4\int_{z_i}^{z_4}dz_3\,\alpha^2\left(z_3\right) 
           \int_{z_i}^{z_3}dz_2\int_{z_i}^{z_2}dz_1\,\alpha^2\left(z_1\right)  \nonumber \\ 
   &   & - \ldots,  
\label{ggenexp}
\end{eqnarray} 
and 
\begin{eqnarray}
h\left(z,z_i\right) 
   & = & \left(z-z_i\right) - \int_{z_i}^{z}dz_2\int_{z_i}^{z_2}dz_1\,\alpha^2
         \left(z_1\right)\left(z_1-z_i\right)  \nonumber \\ 
   &   & + \int_{z_i}^{z}dz_4\int_{z_i}^{z_4}dz_3\,\alpha^2\left(z_3\right) 
           \int_{z_i}^{z_3}dz_2\int_{z_i}^{z_2}dz_1\,\alpha^2\left(z_1\right)\left(z_1-z_i\right)  \nonumber \\ 
   &   & - \ldots.  
\label{hgenexp}
\end{eqnarray} 
Note that the solutions defined by (\ref{ggenexp}) and (\ref{hgenexp}) satisfy automatically 
the required initial conditions (\ref{ghinicondns}).  

\section{Paraxial quantum propagator}  
The $z$-evolution of the wave function$\psi(\vec{r}_\perp,z)$ from the vertical plane at $z_i$ 
on the axis to the plane at $z$ is given by (\ref{psizi2psiz}).  This equation can be written 
in integral form as 
\begin{equation} 
\psi\left(\vec{r}_\perp,z\right) 
   = \int d^2r_{i\perp}\,K\left(\vec{r}_\perp,z;\vec{r}_{\perp i},z_i\right)
     \psi\left(\vec{r}_{\perp i},z_i\right),   
\end{equation} 
where $\int d^2r_{\perp i}$ stands for 
$\int_{-\infty}^{\infty}\int_{-\infty}^{\infty}dx_idy_i$, and 
\begin{equation}  
K\left(\vec{r}_\perp,z;\vec{r}_{\perp i},z_i\right)   
   = \left\langle\vec{r}_\perp\left|\hat{U}\left(z,z_i\right)\right|
                 \vec{r}_{\perp i}\right\rangle, 
\end{equation} 
is the quantum electron beam optical propagator given by the matrix element of the $z$-evolution operator $\hat{U}\left(z,z_i\right)$ in the position representation.  When the beam is paraxial 
the $z$-evolution of $\psi(\vec{r}_\perp,z)$ is given by (\ref{paraxevolneq}).  Then, the 
corresponding paraxial quantum electron beam optical propagator becomes 
\begin{eqnarray}  
K_p\left(\vec{r}_\perp,z;\vec{r}_{\perp i},z_i\right)  
   & = & \left\langle\vec{r}_\perp\left|\hat{U}_p\left(z,z_i\right) 
         \right|\vec{r}_{\perp i}\right\rangle  \nonumber \\ 
   & = & \left\langle\vec{r}_\perp\left|\mathsf{P}\left[\exp{\left(-\frac{i}{\hbar}
         \int_{z_i}^{z}dz\,\hat{\mathcal{H}}_{o,p}(z)\right)}\right] 
         \right|\vec{r}_{\perp i}\right\rangle.  
\end{eqnarray} 
The paraxial quantum electron beam optical Hamiltonian $\hat{\mathcal{H}}_{o,p}$ being quadratic 
in $\left(\vec{r}_\perp,\vec{\hat{p}}_\perp\right)$,  
$K_p\left(\vec{r}_\perp,z;\vec{r}_{\perp i},z_i\right)$ can be calculated exactly 
(see \cite{Wolf1981, Jagannathan1996, Khan1997, Jagannathan2019} 
for details).  The result is 
\begin{eqnarray}
K_p\left(\vec{r}_\perp,z;\vec{r}_{\perp i},z_i\right)  
   & = & \frac{p_0\exp{\left(\frac{i}{\hbar}p_0(z-z_i)\right)}}{i2\pi\hbar h(z,z_i)}  
         \nonumber \\ 
   &   & \quad \times 
   	     \exp{\left[\frac{ip_0}{2\hbar h(z,z_i)}\left(h^\prime(z,z_i)r_\perp^2 
   	     - 2\vec{R}_\perp\cdot\vec{r}_{\perp i} + g(z,z_i)r^2_{\perp i}\right)\right]}, 
         \nonumber \\ 
   &   & \qquad\qquad\qquad\qquad\qquad\qquad \mbox{if}\ h(z,z_i) \neq 0,  
\label{propagatorhneq0}\\ 
K_p\left(\vec{r}_\perp,z;\vec{r}_{\perp i},z_i\right)    
   & = & \frac{1}{g(z,z_i)}\exp{\left[\frac{i}{\hbar}p_0\left((z-z_i) 
   		+ \frac{g^\prime (z,z_i)}{2g(z,z_i)}r_\perp^2\right)\right]}  \nonumber \\ 
   &   & \quad \times\delta^2\left(\frac{\vec{R}_\perp}{g(z,z_i)} - \vec{r}_{\perp i}\right),   
         \quad\ \mbox{if}\ h(z,z_i) = 0,  
\label{propagatorheq0}
\end{eqnarray}
where $\vec{R}_\perp = (X,Y)$ are the coordinates in the rotating coordinate system given by (\ref{xy2XY}).  Thus, for a paraxial beam $z$-evolution of the wave function is given by 
\begin{eqnarray}
\psi\left(\vec{r}_\perp,z\right)  
   & = & \frac{p_0}{i2\pi\hbar h(z,z_i)}
         \exp{\left[\frac{ip_0}{\hbar}\left((z-z_i)+\frac{h^\prime(z,z_i)}
         	        {2h(z,z_i)}r_\perp^2\right)\right]}  \nonumber \\ 
   &   & \quad \times\int d^2r_{\perp i}\exp{\left[\frac{ip_0}{2\hbar h(z,z_i)}     
   	     \left(g(z,z_i)r_{\perp i}^2 - 2\vec{R}_\perp\cdot\vec{r}_{\perp i}\right)\right]}   
         \psi\left(\vec{r}_{\perp i},z_i\right),  \nonumber \\ 
   &   & \qquad\qquad\qquad\qquad\qquad\qquad\qquad\qquad \mbox{if}\ h(z,z_i) \neq 0 
\label{zizparaxpsi1} 
\end{eqnarray}
and 
\begin{eqnarray}
\psi\left(\vec{r}_\perp,z\right)  
   & = & \frac{1}{g(z,z_i)}\exp{\left[\frac{i}{\hbar}p_0\left((z-z_i) 
         + \frac{g^\prime (z,z_i)}{2g(z,z_i)}r_\perp^2\right)\right]} 
           \psi\left(\frac{\vec{R}_\perp}{g(z,z_i)},z_i\right),  \nonumber \\    
   &   & \qquad\qquad\qquad\qquad\qquad\qquad\qquad\qquad \mbox{if}\ h(z,z_i) = 0.  
\label{zizparaxpsi2} 
\end{eqnarray} 
As is well known, equations (\ref{zizparaxpsi1}) and (\ref{zizparaxpsi2}) represent the general 
law of propagation of a paraxial wave function in the case of a round magnetic lens 
\cite{Hawkes1994, Glaser1952, Glaser1953, Glaser1956} and form the basis for the development of 
Fourier transform methods in the electron optical imaging techniques (see \cite{Hawkes1994} for details).  Note that, as follows from (\ref{Mzzi}), if $h\left(z,z_i\right) = 0$ for a 
particular $z$ then $\langle R\rangle(z)$ is the image point of 
$\langle R\rangle\left(z_i\right)$ and the situation corresponds to the stigmatic, or 
point-to-point, imaging with Larmor rotation.  We shall not go further into the well known theory 
of imaging by electron lenses (see \cite{Hawkes1994, Jagannathan1996, Jagannathan2019} and 
references therein for more details).  
  
\section{Glaser model round magnetic lens} 
For the round magnetic lens with Glaser's bell-shaped model for $B(z)$ given by 
(\ref{Glaser}) we have 
\begin{equation}
\alpha(z) = \frac{\alpha_0}{1 + (z/a)^2}, \quad \alpha_0 = \frac{qB_0}{2p_0}.  
\end{equation} 
The corresponding paraxial equation is 
\begin{equation}
\frac{d^2\langle R\rangle(z)}{dz^2}
   + \frac{\alpha_0^2}{\left(1 + (z/a)^2\right)^2}\langle R\rangle(z) = 0,   	
\label{Glaserparaxeq}
\end{equation}
in which $R$ denotes $X$ or $Y$ in the rotating Larmor frame of reference.  Following Glaser 
(see {\em e.g.}, \cite{Hawkes1989b}), let us take 
\begin{equation}
z = a\cot\varphi, \quad \langle R\rangle(\varphi) = \frac{au(\varphi)}{\sin\varphi}.   
\end{equation} 
When $z$ varies from $-\infty$ via $0$ to $\infty$, $\varphi$ varies from $\pi$ via $\pi/2$ to 
$0$.  Then, the equation (\ref{Glaserparaxeq}) becomes 
\begin{equation}
\frac{d^2u(\varphi)}{d\varphi^2} + \omega^2u(\varphi) = 0, \quad 
\mbox{with}\ \omega = \sqrt{1 + \alpha_0^2a^2}.  
\label{phieq}
\end{equation} 
Note that $\omega$ is a constant and hence the general solution of (\ref{phieq}) is given by 
\begin{equation}
u(\varphi) = C\sin[\omega(\varphi-\gamma)], 
\end{equation} 
where $C$ and $\gamma$ are arbitrary constants of integration which have to be fixed to satisfy 
the initial conditions.  Thus, the general solution of the paraxial equation 
(\ref{Glaserparaxeq}) becomes 
\begin{equation}
\langle R\rangle(\varphi) = \frac{Ca\sin[\omega(\varphi - \gamma)]}{\sin\varphi},   
\end{equation} 
in terms of $\varphi$.  Let the electron move from an initial point in the plane at $z = z_i$ 
to the right to the plane at $z > z_i$.  As is well known (see {\em e.g.}, \cite{Lenz1982}), 
the two fundamental solutions of the paraxial equation (\ref{Glaserparaxeq}), in terms of 
$\varphi$ and $\varphi_i$, are given by    
\begin{eqnarray}
g\left(\varphi,\varphi_i\right) 
   & = &  \frac{1}{\omega\sin\varphi}
          \left(\omega\sin\varphi_i\cos\left[\omega\left(\varphi - \varphi_i\right)\right]
          + \cos\varphi_i\sin\left[\omega\left(\varphi - \varphi_i\right)\right]\right)  
          \nonumber \\ 
h\left(\varphi,\varphi_i\right) 
   & = & -\frac{a\sin\left[\omega\left(\varphi - \varphi_i\right)\right]}
         {\omega\sin\varphi_i\sin\varphi}.  
\label{Glasergh}
\end{eqnarray} 
Substituting $\varphi = \cot^{-1}(z/a)$ and $\varphi_i = \cot^{-1}\left(z_i/a\right)$ we get  
$g(z,z_i)$ and $h(z,z_i)$ satisfying the required initial conditions: $g(z_i,z_i) = 1$, 
$g^\prime(z_i,z_i) = 0$, $h(z_i,z_i) = 0$, $h^\prime(z_i,z_i) = 1$.  Then, the `classical' 
trajectory of $\langle\vec{r}_\perp\rangle(z)$, in the laboratory frame of reference is given 
by (\ref{Rpath}).  

Properties of the Glaser model magnetic lens have been well studied since the early days of 
electron microscopy.  We just wanted to show how the classical paraxial theory of such a lens 
follows from quantum mechanics by identifying the classical ray coordinates with the quantum 
averages of position and momentum and using the paraxial approximation at the quantum level. 
The quantum paraxial electron beam optical propagator for the Glaser model lens is obtained 
by substituting the two fundamental solutions (\ref{Glasergh}) in (\ref{zizparaxpsi1}) assuming 
that $h(z,z_i) \neq 0$ for the plane of observation at $z$.  The resulting exact propagator, 
under the paraxial approximation, could be useful in the study of propagation of electron vortex 
beams in a round magnetic lens (see {\em e.g.}, \cite{Loffler2020, Zou2020, Bliokh2017}, and 
references therein, for recent work on related topics).   

\section{Round magnetic lenses with $B(z) \propto z^n$}
Let us now consider the power law model magnetic lenses for which  
\begin{eqnarray}  
B(z) = B_0k_nz^n,                
\end{eqnarray} 
where $n$ is a nonnegative or negative integer.  The case $n = 0$ corresponds to a constant 
magnetic field in the $z$-direction which cannot be used to focus an electron beam.  In this 
case we know the exact solution in the classical situation: helical motion of the electron 
with cyclotron frequency around the $z$-direction. In the quantum situation the eigenstates 
correspond to the Landau levels and one can construct coherent state wave packets for which 
$\langle x\rangle$ and $\langle y\rangle$ follow the classical helical paths \cite{Malkin1969}. 
The field corresponding to $n = -1$ also cannot be used to focus an electron beam 
\cite{Crewe2001}.  So, we shall consider $n$ to be an integer $> 0$ or $< -1$.    

For any integer $n$, $> 0$ or $< -1$, we have 
\begin{equation}  
\alpha(z) = \alpha_0k_nz^n, \quad \mbox{with}\ \alpha_0 = \frac{qB_0}{2p_0}.     
\end{equation} 
The corresponding paraxial equation of motion is 
\begin{equation}  
\frac{d^2\langle R\rangle(z)}{dz^2} + \alpha_0^2k_n^2z^{2n}\langle R\rangle(z) = 0,   
\label{znparaxEq}           
\end{equation} 
where, as earlier, $R$ denotes $X$ or $Y$ in the rotating Larmor frame of reference.  With 
$\alpha_0$ having the dimension of $(\mbox{length})^{-1}$ and $k_nz^n$ being dimensionless 
for both positive and negative values of the integer $n$, note that $\alpha_0k_nz^{n+1}$ is 
dimensionless. Then, defining 
\begin{equation}  
\langle R\rangle(z) = \sqrt{z}\,u(z), 
\end{equation}
and introducing the dimensionless variable 
\begin{equation} 
\zeta = \frac{\alpha_0k_nz^{n+1}}{n+1},  
\end{equation}
the paraxial equation (\ref{znparaxEq}) becomes  
\begin{equation}  
\zeta^2\frac{d^2u(\zeta)}{d\zeta^2} + \zeta\frac{du(\zeta)}{d\zeta}
    + \left(\zeta^2 - \frac{1}{4(n + 1)^2}\right)u(\zeta) = 0.   
\label{zetaeq}
\end{equation} 
Recall that $n$ is not $0$ or $-1$ for us.  Equation (\ref{zetaeq}) is the Bessel differential 
equation of order $1/2(n+1)$ and the two linearly independent solutions for it are the Bessel 
functions of the first kind $J_{1/2(n+1)}(\zeta)$ and $J_{-1/2(n+1)}(\zeta)$.  Bessel function 
of the first kind for any fractional order $\nu$ is defined by  
\begin{eqnarray} 
J_\nu(\zeta) 
   = \frac{1}{\Gamma(\nu+1)}\left(\frac{\zeta}{2}\right)^\nu 
     \sum_{j=0}^{\infty} \frac{(-1)^j}{(\nu + 1)_jj!}\left(\frac{\zeta}{2}\right)^{2j},   
\label{Bessel}
\end{eqnarray} 
where $(a)_j$ is the Pochhammer symbol, or the rising factorial, 
\begin{equation}
(a)_j = \left\{\begin{array}{ll} 
               1, & \mbox{for}\ j = 0, \\ 
               a(a + 1)(a + 2) \ldots (a + j - 1), & \mbox{for}\ j \geq 1,    
               \end{array}\right.,    
\end{equation} 
and $\Gamma$ is the gamma function (for more details on Bessel functions see {\em e.g.}, \cite{Gradshteyn2007, Abramowitz2014, Arfken2012, Lakshminarayanan2015}).  In our case 
$\nu = \pm 1/2(n+1)$.  Thus, the fundamental solutions of the paraxial equation 
(\ref{znparaxEq}) for any integer $n > 0$ are given by 
\begin{eqnarray}
g_n\left(z,z_i\right) 
   & = & C^{(g)}_{n,1}\sqrt{z}J_{\frac{1}{2(n+1)}}
         \left(\frac{\alpha_0k_nz^{n+1}}{n+1}\right) 
         + C^{(g)}_{n,2}\sqrt{z}J_{-\frac{1}{2(n+1)}}
         \left(\frac{\alpha_0k_nz^{n+1}}{n+1}\right),  \nonumber \\ 
h_n\left(z,z_i\right) 
   & = & C^{(h)}_{n,1}\sqrt{z}J_{\frac{1}{2(n+1)}}
         \left(\frac{\alpha_0k_nz^{n+1}}{n+1}\right) 
         + C^{(h)}_{n,2}\sqrt{z}J_{-\frac{1}{2(n+1)}}
         \left(\frac{\alpha_0k_nz^{n+1}}{n+1}\right),   
\label{positiven}
\end{eqnarray} 
where the constants $C^{(g)}_{n,1}$, $C^{(g)}_{n,2}$, $C^{(h)}_{n,1}$, and $C^{(h)}_{n,2}$ are 
to be chosen such that the initial conditions (\ref{ghinicondns}) are satisfied.  Note that when 
$B(z) = B_0k_{-n}z^{-n}$ we have $\zeta = -\alpha_0k_{-n}/((n-1)z^{n-1})$ and in (\ref{zetaeq}) 
we can write $1/(2(-n+1))^2$ as $1/(2(n-1))^2$.  Note also that 
$J_\nu(-\zeta) = (-1)^\nu J_\nu(\zeta)$.  As a result, the corresponding fundamental solutions 
of the paraxial equation (\ref{znparaxEq}) for any integer $n < -1$ can be written as 
\begin{eqnarray}
g_{-n}\left(z,z_i\right) 
   & = & C^{(g)}_{-n,1}\sqrt{z}J_{\frac{1}{2(n-1)}}
         \left(\frac{\alpha_0k_{-n}}{(n-1)z^{n-1}}\right)  
         + C^{(g)}_{-n,2}\sqrt{z}J_{-\frac{1}{2(n-1)}}
         \left(\frac{\alpha_0k_{-n}}{(n-1)z^{n-1}}\right), \nonumber \\
h_{-n}\left(z,z_i\right) 
   & = & C^{(h)}_{-n,1}\sqrt{z}J_{\frac{1}{2(n-1)}}
         \left(\frac{\alpha_0k_{-n}}{(n-1)z^{n-1}}\right) 
         + C^{(h)}_{-n,2}\sqrt{z}J_{-\frac{1}{2(n-1)}}
         \left(\frac{\alpha_0k_{-n}}{(n-1)z^{n-1}}\right),  \nonumber \\ 
   &   &   
\label{negativen}
\end{eqnarray}
where the constants $C^{(g)}_{-n,1}$, $C^{(g)}_{-n,2}$, $C^{(h)}_{-n,1}$, and $C^{(h)}_{-n,2}$ 
are to be chosen such that the initial conditions (\ref{ghinicondns}) are satisfied.  We shall 
look at some examples in the following.  

For any integer $n > 0$, we shall take $z_i = 0$ so that the electron moves from the field-free 
region to the field region on the right ($z > 0$).  Then the initial conditions to be satisfied 
are: 
\begin{equation} 
g_n(0,0) = 1, \quad g_n^\prime(0,0) = 0,  \qquad 
h_n(0,0) = 0, \quad h_n^\prime(0,0) = 1. 
\end{equation} 
From (\ref{Bessel}) it is observed that  
\begin{equation} 
\sqrt{z}J_{\frac{1}{2(n+1)}}\left(\frac{\alpha_0k_nz^{n+1}}{n+1}\right) = 0, 
\quad \mbox{at}\ z = 0,     
\end{equation} 
and 
\begin{equation}
\sqrt{z}J_{-\frac{1}{2(n+1)}}\left(\frac{\alpha_0k_nz^{n+1}}{n+1}\right)   
   = \frac{1}{\Gamma\left(\frac{2n+1}{2n+2}\right)} 
   	 \left(\frac{\alpha_0k_n}{2(n+1)}\right)^{-\frac{1}{2(n+1)}}, \quad \mbox{at}\ z = 0.	
\end{equation}
This suggests that we can take  
\begin{equation} 
C^{(g)}_{n,1} = 0,   
C^{(g)}_{n,2} = \Gamma\left(\frac{2n+1}{2n+2}\right)
                \left(\frac{\alpha_0k_n}{2(n+1)}\right)^{\frac{1}{2(n+1)}},  
\end{equation} 
so that $g_n(0,0) = 1$.  In other words, 
\begin{eqnarray} 
g_n(z,0) = \Gamma\left(\frac{2n+1}{2(n+1)}\right)
           \left(\frac{\alpha_0k_n}{2(n+1)}\right)^{\frac{1}{2(n+1)}} 
           \left[\sqrt{z}J_{-\frac{1}{2(n+1)}}
           \left(\frac{\alpha_0k_nz^{n+1}}{n+1}\right)\right].  
\label{gnz0}
\end{eqnarray}
It is seen from (\ref{Bessel}) that the first term of $g_n(z,0)$ is a constant and from the 
second term onwards, the $j$-th term is $\propto z^{2(j-1)(n+1)}$.  From this it follows that $g_n^\prime(0,0) = 0$ as required.  Similar arguments lead to the result: 
\begin{eqnarray} 
h_n(z,0) 
   & = & \Gamma\left(\frac{2n+3}{2(n+1)}\right)
         \left(\frac{\alpha_0k_n}{2(n+1)}\right)^{-\frac{1}{2(n+1)}}  
         \left[\sqrt{z}J_{\frac{1}{2(n+1)}}\left(\frac{\alpha_0k_nz^{n+1}}{n+1}\right)\right]. 
         \nonumber \\ 
   &   &  
\label{hnz0}
\end{eqnarray}
Using (\ref{Bessel}) it can be verified that $h_n(0,0) = 0$ and $h_n^\prime(0,0) = 1$.  

For any integer $n < -1$ the field vanishes as $z \longrightarrow \pm\infty$.  We shall 
consider the electron to move from the field-free region at $-\infty$ towards the field region 
on the right {\em i.e.}, $z_i = -\infty$.  Then the initial conditions are: 
$\lim_{z\rightarrow -\infty} g_{-n}\left(z,-\infty\right) = 1$, 
$\lim_{z\rightarrow -\infty}g_{-n}^\prime\left(z,-\infty\right) = 0$,  
$\lim_{z\rightarrow -\infty}h_{-n}\left(z,-\infty\right) = 0$, 
$\lim_{z\rightarrow -\infty}h_{-n}^\prime\left(z,-\infty\right) = 1$.  
From (\ref{Bessel}) it is observed that    
\begin{equation}
\lim_{z\rightarrow -\infty}\sqrt{z}J_{\frac{1}{2(n-1)}}
       \left(\frac{\alpha_0k_{-n}}{(n-1)z^{n-1}}\right)  
\longrightarrow \frac{1}{\Gamma\left(\frac{2n-1}{2(n-1)}\right)} 
                \left(\frac{\alpha_0k_{-n}}{2(n-1)}\right)^{\frac{1}{2(n-1)}} 	
\label{inftylim1}
\end{equation} 
and 
\begin{equation} 
\lim_{z\rightarrow -\infty}\sqrt{z}J_{-\frac{1}{2(n-1)}}
       \left(\frac{\alpha_0k_{-n}}{(n-1)z^{n-1}}\right)  
\longrightarrow \frac{1}{\Gamma\left(\frac{2n-3}{2(n-1)}\right)} 
                \left(\frac{\alpha_0k_{-n}}{2(n-1)}\right)^{-\frac{1}{2(n-1)}}\,z.      
\label{inftylim2}
\end{equation}    
Then, it follows that we should take 
\begin{equation}
C^{(g)}_{-n,1} = \Gamma\left(\frac{2n-1}{2(n-1)}\right)
                 \left(\frac{\alpha_0k_{-n}}{2(n-1)}\right)^{-\frac{1}{2(n-1)}}, \quad 
C^{(g)}_{-n,2} = 0,      	
\end{equation} 
so that $\lim_{z\rightarrow -\infty}g_{-n}\left(z,-\infty\right) = 1$.  In other words, 
\begin{equation} 
g_{-n}(z,-\infty)   
   = \Gamma\left(\frac{2n-1}{2(n-1)}\right)
     \left(\frac{\alpha_0k_{-n}}{2(n-1)}\right)^{-\frac{1}{2(n-1)}}  
     \left[\sqrt{z}J_{\frac{1}{2(n-1)}}
     \left(\frac{\alpha_0k_{-n}}{(n-1)z^{n-1}}\right)\right].  
\label{g-nzinfty} 
\end{equation}
It is seen from (\ref{Bessel}) that the first term of $g_{-n}(z,-\infty)$ is a constant and 
from the second term onwards, the $j$-th term is $\propto z^{-2(j-1)(n-1)}$.  From this it 
follows that $\lim_{z\rightarrow -\infty}g^\prime(z,\infty) = 0$ as required.  In view of the 
asymptotic limits in 
(\ref{inftylim1}) and (\ref{inftylim2}) we can take 
\begin{eqnarray}
h_{-n}(z,-\infty)  
   & = & \left\{\Gamma\left(\frac{2n-3}{2(n-1)}\right)
         \left(\frac{\alpha_0k_{-n}}{2(n-1)}\right)^{\frac{1}{2(n-1)}} \right. \nonumber \\ 
   &   & \qquad\qquad\qquad \times  
         \left[\sqrt{z}J_{-\frac{1}{2(n-1)}}
         \left(\frac{\alpha_0k_{-n}}{(n-1)z^{n-1}}\right)\right]  \nonumber \\ 
   &   & \quad -\lim_{z_i\rightarrow -\infty}z_i\Gamma\left(\frac{2n-1}{2(n-1)}\right)
         \left(\frac{\alpha_0k_{-n}}{2(n-1)}\right)^{-\frac{1}{2(n-1)}} \nonumber \\ 
   &   & \qquad\qquad\qquad \left.\times\left[\sqrt{z}J_{\frac{1}{2(n-1)}}
         \left(\frac{\alpha_0k_{-n}}{(n-1)z^{n-1}}\right)\right]\right\},   	
\label{h-nzinfty}
\end{eqnarray} 
such that $\lim_{z\rightarrow -\infty}h_{-n}\left(z,-\infty\right) = 0$ and  
$\lim_{z\rightarrow -\infty}h_{-n}^\prime\left(z,-\infty\right) = 1$. 

Let us now see how the Peano-Baker method works.  From the general expressions in 
(\ref{ggenexp}) and (\ref{hgenexp}) we have  
\begin{eqnarray} 
g_n\left(z,z_i\right)  
   & = & 1 - \left(\alpha_0k_n\right)^2\int_{z_i}^{z}dz_2
                                       \int_{z_i}^{z_2}dz_1\,z_1^{2n}  \nonumber \\   
   &   & + \left(\alpha_0k_n\right)^4\int_{z_i}^{z}dz_4
                            \int_{z_i}^{z_4}dz_3\,z_3^{2n}\int_{z_i}^{z_3}dz_2
                            \int_{z_i}^{z_2}dz_1\,z_1^{2n} \nonumber \\ 
   &   & - \ldots   
\label{ggenexp4n}
\end{eqnarray} 
and 
\begin{eqnarray}  
h_n\left(z,z_i\right)   
   & = &  \left(z - z_i\right)   
        - \left(\alpha_0k_n\right)^2\int_{z_i}^{z}dz_2
          \int_{z_i}^{z_2}dz_1\,z_1^{2n}\left(z_1-z_i\right)  \nonumber \\   
   &   & + \left(\alpha_0k_n\right)^4\int_{z_i}^{z}dz_4\int_{z_i}^{z_4}dz_3\,z_3^{2n}  
           \int_{z_i}^{z_3}dz_2\int_{z_i}^{z_2}dz_1\,z_1^{2n}\left(z_1-z_i\right)  \nonumber \\ 
   &   & - \ldots,  
\label{hgenexp4n}
\end{eqnarray} 
where the electron is considered to be moving from $z_i$ to $z$.  It should be noted that the 
initial conditions are satisfied automatically.  

For any integer $n > 0$, with $z_i = 0$, it is straightforward to see from (\ref{ggenexp4n}) 
and (\ref{hgenexp4n}) that 
\begin{eqnarray}
g_n(z,0)    
   & = & 1 - \frac{\left(\alpha_0k_nz^{n+1}\right)^2}{(2n+1)(2n+2)}  
         + \frac{\left(\alpha_0k_nz^{n+1}\right)^4}{(2n+1)(2n+2)(4n+3)(4n+4)} 
         - \ldots \nonumber \\ 
   & = & \sum_{j=0}^{\infty} \frac{(-1)^j}{\left(\frac{2n+1}{2(n+1)}\right)_jj!}
                             \left(\frac{\alpha_0k_nz^{n+1}}{2(n+1)}\right)^{2j}  \nonumber \\ 
   & = & \Gamma\left(\frac{2n+1}{2(n+1)}\right)
               \left(\frac{\alpha_0k_n}{2(n+1)}\right)^{\frac{1}{2(n+1)}}  
               \left[\sqrt{z}J_{-\frac{1}{2(n+1)}}
               \left(\frac{\alpha_0k_nz^{n+1}}{n+1}\right)\right]  
\label{g4positiven}
\end{eqnarray} 
and 
\begin{eqnarray} 
h_n(z,0)    
   & = & z\left[1 - \frac{\left(\alpha_0k_nz^{n+1}\right)^2}{(2n+2)(2n+3)}  
         + \frac{\left(\alpha_0k_nz^{n+1}\right)^4}
                                  {(2n+2)(2n+3)(4n+4)(4n+5)} - \ldots\right]   \nonumber \\  
   & = & z\sum_{j=0}^{\infty}\frac{(-1)^j}{\left(\frac{2n+3}{2(n+1)}\right)_jj!}
                              \left(\frac{\alpha_0k_nz^{n+1}}{2(n+1)}\right)^{2j}  \nonumber \\  
   & = & \Gamma\left(\frac{2n+3}{2(n+1)}\right)
               \left(\frac{\alpha_0k_n}{2(n+1)}\right)^{-\frac{1}{2(n+1)}}  
               \left[\sqrt{z}J_{\frac{1}{2(n+1)}}
               \left(\frac{\alpha_0k_nz^{n+1}}{n+1}\right)\right].  
\label{h4positiven}
\end{eqnarray}    
For any integer $n < -1$, with $z_i = -\infty$, we have 
\begin{eqnarray} 
g_{-n}(z,-\infty)  
   & = & 1 - \frac{1}{(2n-1)(2n-2)}\left(\frac{\alpha_0k_{-n}}{z^{n-1}}\right)^2  \nonumber \\ 
   &   & \quad + \frac{1}{(2n-1)(2n-2)(4n-3)(4n-4)} 
                 \left(\frac{\alpha_0k_{-n}}{z^{n-1}}\right)^4 - \ldots \nonumber \\ 
   & = & \sum_{j=0}^{\infty} \frac{(-1)^j}{\left(\frac{2n-1}{2(n-1)}\right)_jj!}
                             \left(\frac{\alpha_0k_{-n}}{2(n-1)z^{n-1}}\right)^{2j}  
                             \nonumber \\ 
   & = & \Gamma\left(\frac{2n-1}{2(n-1)}\right)
         \left(\frac{\alpha_0k_{-n}}{2(n-1)}\right)^{-\frac{1}{2(n-1)}}  \nonumber \\ 
   &   & \qquad\qquad\qquad \times\left[\sqrt{z}J_{\frac{1}{2(n-1)}} 
         \left(\frac{\alpha_0k_{-n}}{(n-1)z^{n-1}}\right)\right]  
\label{g4negativen}
\end{eqnarray} 
and 
\begin{eqnarray}
h_{-n}(z,-\infty)  
   & = & z\left[1 - \frac{1}{(2n-2)(2n-3)}
                    \left(\frac{\alpha_0k_{-n}}{z^{n-1}}\right)^2 \right.  \nonumber \\ 
   &   & \qquad \left. + \frac{1}{(2n-2)(2n-3)(4n-4)(4n-5)} 
                   \left(\frac{\alpha_0k_{-n}}{z^{n-1}}\right)^4 - \ldots\right] \nonumber \\ 
   &   & -\lim_{z_i\rightarrow -\infty}
          \left\{z_i\left[1 - \frac{1}{(2n-1)(2n-2)}
          \left(\frac{\alpha_0k_{-n}}{z^{n-1}}\right)^2  \right.\right. \nonumber \\ 
   &   &  \quad \left.\left. + \frac{1}{(2n-1)(2n-2)(4n-3)(4n-4)} 
          \left(\frac{\alpha_0k_{-n}}{z^{n-1}}\right)^4 - \ldots \right]\right\}  \nonumber \\ 
   & = & z\left\{\sum_{j=0}^{\infty}\frac{(-1)^j}{\left(\frac{2n-3}{2(n-1)}\right)_jj!}
          \left(\frac{\alpha_0k_{-n}}{2(n-1)z^{n-1}}\right)^{2j}\right\}  \nonumber \\ 
   &   & \quad -\lim_{z_i\rightarrow -\infty}
                \left\{z_i\left[\sum_{j=0}^{\infty}
                \frac{(-1)^j}{\left(\frac{2n-1}{2(n-1)}\right)_jj!}
         \left(\frac{\alpha_0k_{-n}}{2(n-1)z^{n-1}}\right)^{2j}\right]\right\}  \nonumber \\ 
   & = & \left\{\Gamma\left(\frac{2n-3}{2(n-1)}\right)
         \left(\frac{\alpha_0k_{-n}}{2(n-1)}\right)^{\frac{1}{2(n-1)}} \right. \nonumber \\ 
   &   & \qquad\qquad\qquad \times \left[\sqrt{z}J_{-\frac{1}{2(n-1)}}
                     \left(\frac{\alpha_0k_{-n}}{(n-1)z^{n-1}}\right)\right]  \nonumber \\ 
   &   & \quad -\lim_{z_i\rightarrow -\infty}z_i\Gamma\left(\frac{2n-1}{2(n-1)}\right)
                \left(\frac{\alpha_0k_{-n}}{2(n-1)}\right)^{-\frac{1}{2(n-1)}}  \nonumber \\ 
   &   & \qquad\qquad\qquad \left. \times\left[\sqrt{z}J_{\frac{1}{2(n-1)}} 
                     \left(\frac{\alpha_0k_{-n}}{(n-1)z^{n-1}}\right)\right]\right\}.    
\label{h4negativen}
\end{eqnarray}        
We see that the solutions (\ref{g4positiven}-\ref{h4negativen}) are the same as those obtained analytically in (\ref{gnz0}), (\ref{hnz0}), (\ref{g-nzinfty}), and (\ref{h-nzinfty}).  Thus, 
the Peano-Baker method is a constructive process, incorporating the initial conditions, leading 
to the same solutions as the analytical solutions.  When it is not possible to get the analytical solutions in a particular case the Peano-Baker method provides a computational scheme to get the  approximate solutions (see {\em e.g.}, \cite{Vaseghi2020}).  

The above analytical solutions are well known and, based on them, practical aspects of the 
performances of axially symmetric magnetic lenses with the Glaser and power law models for 
$B(z)$ have been analysed extensively in the literature (see \cite{Hawkes1989a, Hawkes1989b, Hawkes1982, Gianola1952, Hansel1964, Alshwaikh1977, AlHilly1982, Mulvey1982, Lenc1992, 
Crewe2001, Crewe2003, Liu2003, Crewe2004, Hawkes2002, Alamir2003, Alamir2004, Alamir2005, 
Alamir2009a, Alamir2009b, Alamir2011}).  Here, our objective has been mainly to demonstrate, 
with the examples of some models of round magnetic lenses, how quantum mechanics leads to the 
classical trajectories for the quantum average of position when the small quantum corrections 
are neglected.   

\section{Quantum mechanics of aberrations} 
When the incoming electron beam is not ideally paraxial we have to retain in the quantum 
electron beam optical Hamiltonian $\hat{\mathcal{H}}_{o,c}$ terms higher than quadratic in 
$\langle x\rangle$, $\langle y\rangle$, $\langle\hat{p}_x\rangle$, and 
$\langle\hat{p}_y\rangle$ which lead to terms not linear in $\langle x\rangle$, 
$\langle y\rangle$, $\langle\hat{p}_x\rangle$, and 
$\langle\hat{p}_y\rangle$ in the equations of motion (\ref{EqMx}-\ref{EqMpy}).  Thus, for a quasiparaxial beam we shall take 
\begin{equation} 
\hat{\mathcal{H}}_{o,c} = \hat{\mathcal{H}}_{o,p} + \hat{\mathcal{H}}^\prime_o, 
\end{equation} 
as the Hamiltonian, where $\hat{\mathcal{H}}_{o,p}$ and $\hat{\mathcal{H}}^\prime_o$ are, 
respectively, the paraxial Hamiltonian and the lowest order non-paraxial Hamiltonian given in   (\ref{Hop}-\ref{Hoprime}).  From (\ref{Oz}) we know that when the beam is paraxial the relation 
between $\langle\hat{O}\rangle(z_i)$ and $\langle\hat{O}\rangle(z)$, for any observable $O$, 
is given by 
\begin{equation} 
\langle\hat{O}\rangle(z) 
  = \left\langle\hat{U}_p^\dagger\left(z,z_i\right)\hat{O}\hat{U}_p\left(z,z_i\right)
    \right\rangle\left(z_i\right),    
\label{paraxOz} 
\end{equation} 
where 
\begin{equation}
\hat{U}_p\left(z,z_i\right) 
   = \mathsf{P}\left[\exp{\left(-\frac{i}{\hbar}
   	                 \int_{z_i}^{z}dz\,\hat{\mathcal{H}}_{o,p}(z)\right)}\right].  
\end{equation}    
From (\ref{paraxtrnsfrmap}) we have the paraxial transfer map:  
\begin{eqnarray}
\left(\begin{array}{c}
      \langle x\rangle_p(z) \\  
      \langle y\rangle_p(z) \\ 
      \frac{1}{p_0}\langle\hat{p}_x\rangle_p(z) \\ 
      \frac{1}{p_0}\langle\hat{p}_y\rangle_p(z) 
      \end{array}\right)   
   & = & \left(\begin{array}{c}
               \langle\hat{U}_p^\dagger\left(z,z_i\right)x 
               \hat{U}_p\left(z,z_i\right)\rangle\left(z_i\right) \\ 
               \langle\hat{U}_p^\dagger\left(z,z_i\right)y 
               \hat{U}_p\left(z,z_i\right)\rangle\left(z_i\right) \\ 
               \frac{1}{p_0}\langle\hat{U}_p^\dagger\left(z,z_i\right)\hat{p}_x 
               \hat{U}_p\left(z,z_i\right)\rangle\left(z_i\right) \\ 
               \frac{1}{p_0}\langle\hat{U}_p^\dagger\left(z,z_i\right)\hat{p}_y
               \hat{U}_p\left(z,z_i\right)\rangle\left(z_i\right)            
               \end{array}\right)  \nonumber \\ 
   & = & \left(\begin{array}{cc}
                    g\left(z,z_i\right)\mathsf{R}\left(z,z_i\right) & 
                    h\left(z,z_i\right)\mathsf{R}\left(z,z_i\right) \\ 
                    g^\prime\left(z,z_i\right)\mathsf{R}\left(z,z_i\right) & 
                    h^\prime\left(z,z_i\right)\mathsf{R}\left(z,z_i\right)   
                     \end{array}\right)  \nonumber \\ 
   &   & \quad \times\left(\begin{array}{c}
                           \langle x\rangle\left(z_i\right) \\  
                           \langle y\rangle\left(z_i\right) \\ 
                           \frac{1}{p_0}\langle\hat{p}_x\rangle\left(z_i\right) \\ 
                           \frac{1}{p_0}\langle\hat{p}_y\rangle\left(z_i\right)  
                           \end{array}\right).                   
\label{idealmap}
\end{eqnarray} 
To get the transfer map for the quasiparaxial beam we have to calculate  
\begin{equation}
\left(\begin{array}{c}
      \langle x\rangle(z) \\  
      \langle y\rangle(z) \\ 
      \frac{1}{p_0}\langle\hat{p}_x\rangle(z) \\ 
      \frac{1}{p_0}\langle\hat{p}_y\rangle(z) 
      \end{array}\right)  
   = \left(\begin{array}{c}
           \langle\hat{U}^\dagger\left(z,z_i\right)x 
           \hat{U}\left(z,z_i\right)\rangle\left(z_i\right) \\ 
           \langle\hat{U}^\dagger\left(z,z_i\right)y 
           \hat{U}\left(z,z_i\right)\rangle\left(z_i\right) \\ 
           \frac{1}{p_0}\langle\hat{U}^\dagger\left(z,z_i\right)\hat{p}_x 
           \hat{U}\left(z,z_i\right)\rangle\left(z_i\right) \\ 
           \frac{1}{p_0}\langle\hat{U}^\dagger\left(z,z_i\right)\hat{p}_y
           \hat{U}\left(z,z_i\right)\rangle\left(z_i\right)            
           \end{array}\right),  
\label{nonparaxmap}
\end{equation}
where 
\begin{eqnarray} 
\hat{U}\left(z,z_i\right)  
   & = & \mathsf{P}\left[\exp{\left(-\frac{i}{\hbar}
               	         \int_{z_i}^{z}dz\,\hat{\mathcal{H}}_{o,c}(z)\right)}\right]  
                         \nonumber \\ 
   & = & \mathsf{P}\left[\exp{\left(-\frac{i}{\hbar}
   	                     \int_{z_i}^{z}dz\,\left(\hat{\mathcal{H}}_{o,p}(z) 
   	                   + \hat{\mathcal{H}}^\prime_o(z)\right)\right)}\right].  
\end{eqnarray}
To get the transfer map (\ref{nonparaxmap}), one has to use the formalism of the time-dependent perturbation theory of quantum mechanics based on the interaction picture, replacing time $t$ 
by $z$, treating $\hat{\mathcal{H}}^\prime_o(z)$ as a perturbation.  The results corresponding 
to imaging by a round magnetic lens are as follows (for details see, 
\cite{Jagannathan1996, Jagannathan2019}).  

Let us now choose $z_i = z_{ob}$, the position of the object plane, and $z = z_{im}$, the 
position of the image plane where a magnified, inverted, and rotated, image of the object is 
formed.  Under ideal conditions we would have point-to-point imaging such that in 
(\ref{idealmap}) $g\left(z_{im},z_{ob}\right) = -M$, $h\left(z_{im},z_{ob}\right) = 0$, $g^\prime\left(z_{im},z_{ob}\right) = -1/f$, and $h^\prime\left(z_{im},z_{ob}\right) = -1/M$ 
where $M$ is the magnification and $f$ is the focal length of the lens.  Under quasiparaxial 
conditions the transfer map becomes 
\begin{eqnarray}
\left(\begin{array}{c}
      \langle x\rangle\left(z_{im}\right) \\  
      \langle y\rangle\left(z_{im}\right) \\ 
      \frac{1}{p_0}\langle\hat{p}_x\rangle\left(z_{im}\right) \\ 
      \frac{1}{p_0}\langle\hat{p}_y\rangle\left(z_{im}\right) 
      \end{array}\right)   
   & = & \left(\begin{array}{c}
               \langle x\rangle_p\left(z_{im}\right) \\  
               \langle y\rangle_p\left(z_{im}\right) \\ 
               \frac{1}{p_0}\langle\hat{p}_x\rangle_p\left(z_{im}\right) \\ 
               \frac{1}{p_0}\langle\hat{p}_y\rangle_p\left(z_{im}\right) 
               \end{array}\right) 
                + \left(\begin{array}{c}
                       (\Delta x)\left(z_{im}\right) \\ 
                       (\Delta y)\left(z_{im}\right) \\ 
                        \frac{1}{p_0}\left(\Delta p_x\right)\left(z_{im}\right) \\ 
                        \frac{1}{p_0}\left(\Delta p_y\right)\left(z_{im}\right)
                        \end{array}\right)  \nonumber \\        
   & = & \left(\begin{array}{cc}
                     -M\mathsf{R}\left(z_{im},z_{ob}\right) & \mathbf{0} \\
                      -\frac{1}{f}\mathsf{R}\left(z_{im},z_{ob}\right) & 
                      -\frac{1}{M}\mathsf{R}\left(z_{im},z_{ob}\right)   
                       \end{array}\right)  \nonumber \\ 
   &   & \quad \times\left(\left(\begin{array}{c}
                                 \langle x\rangle\left(z_{ob}\right) \\  
                                 \langle y\rangle\left(z_{ob}\right) \\ 
                                 \frac{1}{p_0}\langle\hat{p}_x\rangle\left(z_{ob}\right) \\ 
                                 \frac{1}{p_0}\langle\hat{p}_y\rangle\left(z_{ob}\right)  
                                 \end{array}\right) + 
                      \left(\begin{array}{c}
                           (\delta x)\left(z_{ob}\right) \\ 
                           (\delta y)\left(z_{ob}\right) \\ 
                            \frac{1}{p_0}\left(\delta p_x\right)\left(z_{ob}\right)\\ 
                            \frac{1}{p_0}\left(\delta p_y\right)\left(z_{ob}\right)
                            \end{array}\right)\right), \nonumber \\ 
   &   &  
\end{eqnarray}
where $(\Delta x)(z_{im})$, $(\Delta y)(z_{im})$, $\left(\Delta p_x\right)(z_{im})$, and 
$\left(\Delta p_y\right)(z_{im})$ are the third-order aberrations, lowest order deviations 
from the paraxial results involving the quantum averages of third-order polynomials in $\left(\vec{r}_\perp,\vec{\hat{p}}_\perp\right)$ as seen in the following.  Explicit 
expressions for $(\delta x)\left(z_{ob}\right)$, $(\delta y)\left(z_{ob}\right)$, 
$\left(\delta p_x\right)\left(z_{ob}\right)$, and $\left(\delta p_y\right)\left(z_{ob}\right)$ 
are given by   
\begin{eqnarray} 
(\delta x)\left(z_{ob}\right) 
   & = & \frac{C}{p_0^3}\left\langle\hat{p}_x\hat{p}_\perp^2\right\rangle\left(z_{ob}\right)  
         \nonumber \\   
   &   & + \frac{K}{2p_0^2}\left\langle\left\{\hat{p}_x,\left(\vec{\hat{p}}_\perp\cdot
                 \vec{r}_\perp + \vec{r}_\perp\cdot\vec{\hat{p}}_\perp\right)\right\}   
               + \left\{x,\hat{p}_\perp^2\right\}\right\rangle\left(z_{ob}\right)  \nonumber \\  
   &   & + \frac{k}{p_0^2}\left\langle\left\{\hat{p}_x,\hat{L}_z\right\} 
               - \frac{1}{2}\left\{y,\hat{p}_\perp^2\right\}\right\rangle\left(z_{ob}\right) 
                 \nonumber \\  
   &   & + \frac{A}{2p_0}\left\langle\left\{x,\left(\vec{\hat{p}}_\perp\cdot 
                 \vec{r}_\perp+\vec{r}_\perp\cdot\vec{\hat{p}}_\perp\right)\right\}\right\rangle
                 \left(z_{ob}\right)  \nonumber \\ 
   &   & + \frac{a}{2p_0}\left\langle\left\{x,\hat{L}_z\right\}   
               - \left\{y,\left(\vec{\hat{p}}_\perp\cdot\vec{r}_\perp+\vec{r}_\perp\cdot
                 \vec{\hat{p}}_\perp\right)\right\}\right\rangle\left(z_{ob}\right)  \nonumber \\  
   &   & + \frac{F}{2p_0}\left\langle\left\{\hat{p}_x,r_\perp^2\right\}
                 \right\rangle\left(z_{ob}\right)    
         + D\left\langle xr_\perp^2\right\rangle\left(z_{ob}\right) 
         - d\left\langle yr_\perp^2\right\rangle\left(z_{ob}\right),  
\label{xaberration} \\ 
(\delta y)\left(z_{ob}\right) 
   & = & \frac{C}{p_0^3}\left\langle\hat{p}_y\hat{p}_\perp^2\right\rangle\left(z_{ob}\right)  
         \nonumber \\ 
   &   & + \frac{K}{2p_0^2}\left\langle\left\{\hat{p}_y,\left(\vec{\hat{p}}_\perp\cdot
                 \vec{r}_\perp+\vec{r}_\perp\cdot\vec{\hat{p}}_\perp\right)\right\} 
               + \left\{y,\hat{p}_\perp^2\right\}\right\rangle\left(z_{ob}\right)  \nonumber \\ 
   &   & + \frac{k}{p_0^2}\left\langle\left\{\hat{p}_y,\hat{L}_z\right\}
               + \frac{1}{2}\left\{x,\hat{p}_\perp^2\right\}\right\rangle\left(z_{ob}\right)  
                 \nonumber \\ 
   &   & + \frac{A}{2p_0}\left\langle\left\{y,\left(\vec{\hat{p}}_\perp\cdot
                 \vec{r}_\perp+\vec{r}_\perp\cdot\vec{\hat{p}}_\perp\right)\right\}\right\rangle
                 \left(z_{ob}\right) \nonumber \\ 
   &   & + \frac{a}{2p_0}\left\langle\left\{y,\hat{L}_z\right\} 
               - \left\{x,\left(\vec{\hat{p}}_\perp\cdot\vec{r}_\perp+\vec{r}_\perp\cdot
                 \vec{\hat{p}}_\perp\right)\right\}\right\rangle\left(z_{ob}\right)  \nonumber \\   
   &   & + \frac{F}{2p_0}\left\langle\left\{\hat{p}_y,r_\perp^2\right\}
                 \right\rangle\left(z_{ob}\right)   
         + D\left\langle yr_\perp^2\right\rangle\left(z_{ob}\right)   
         + d\left\langle xr_\perp^2\right\rangle\left(z_{ob}\right),  
\label{yaberration} \\ 
\frac{1}{p_0}(\delta p_x)\left(z_{ob}\right)     
   & = & -\frac{K}{p_0^3}\left\langle\hat{p}_x\hat{p}_\perp^2\right\rangle\left(z_{ob}\right) 
         - \frac{k}{p_0^3}\left\langle\hat{p}_y\hat{p}_\perp^2\right\rangle\left(z_{ob}\right)      
           \nonumber \\ 
   &   & - \frac{A}{2p_0^2}\left\langle\left\{\hat{p}_x,\left(\vec{\hat{p}}_\perp\cdot
                 \vec{r}_\perp + \vec{r}_\perp\cdot\vec{\hat{p}}_\perp\right)\right\}
                 \right\rangle\left(z_{ob}\right)  \nonumber \\ 
   &   & - \frac{a}{2p_0^2}\left\langle\left\{\hat{p}_x,\hat{L}_z\right\} 
               + \left\{\hat{p}_y,\left(\vec{\hat{p}}_\perp\cdot\vec{r}_\perp 
               + \vec{r}_\perp\cdot\vec{\hat{p}}_\perp\right)\right\}\right\rangle
                 \left(z_{ob}\right)  \nonumber \\    
   &   & - \frac{F}{2p_0^2}\left\langle\left\{x,\hat{p}_\perp^2\right\}
                 \right\rangle\left(z_{ob}\right)  \nonumber \\   
   &   & - \frac{D}{2p_0}\left\langle\left\{\hat{p}_x,r_\perp^2\right\}  
               + \left\{x,\left(\vec{\hat{p}}_\perp\cdot\vec{r}_\perp 
               + \vec{r}_\perp\cdot\vec{\hat{p}}_\perp\right)\right\}\right\rangle
                 \left(z_{ob}\right)  \nonumber \\ 
   &   & - \frac{d}{p_0}\left\langle\left\{x,\hat{L}_z\right\} 
               + \frac{1}{2}\left\{\hat{p}_y,r_\perp^2\right\}\right\rangle\left(z_{ob}\right) 
        - E\left\langle xr_\perp^2\right\rangle\left(z_{ob}\right),  
\label{pxaberration} \\ 
\frac{1}{p_0}(\delta p_y)\left(z_{ob}\right)  
   & = & - \frac{K}{p_0^3}\left\langle\hat{p}_y\hat{p}_\perp^2\right\rangle\left(z_{ob}\right) 
         + \frac{k}{p_0^3}\left\langle\hat{p}_x\hat{p}_\perp^2\right\rangle\left(z_{ob}\right)     
           \nonumber \\  
   &   & - \frac{A}{2p_0^2}\left\langle\left\{\hat{p}_y,\left(\vec{\hat{p}}_\perp\cdot
           \vec{r}_\perp+\vec{r}_\perp\cdot\vec{\hat{p}}_\perp\right)\right\}
           \right\rangle\left(z_{ob}\right)  \nonumber \\ 
   &   & - \frac{a}{2p_0^2}\left\langle\left\{\hat{p}_y,\hat{L}_z\right\} 
         + \left\{\hat{p}_x,\left(\vec{\hat{p}}_\perp\cdot 
           \vec{r}_\perp + \vec{r}_\perp\cdot\vec{\hat{p}}_\perp\right)\right\}
           \right\rangle\left(z_{ob}\right)  \nonumber \\   
   &   & - \frac{F}{2p_0^2}\left\langle\left\{y,\hat{p}_\perp^2\right\}
           \right\rangle\left(z_{ob}\right)  \nonumber \\  
   &   & - \frac{D}{2p_0}\left\langle\left\{\hat{p}_y,r_\perp^2\right\} 
         + \left\{y,\left(\vec{\hat{p}}_\perp\cdot\vec{r}_\perp 
         + \vec{r}_\perp\cdot\vec{\hat{p}}_\perp\right)\right\}\right\rangle\left(z_{ob}\right)         
           \nonumber \\ 
   &   & - \frac{d}{p_0}\left\langle\left\{y,\hat{L}_z\right\}-\frac{1}{2}
           \left\{\hat{p}_x,r_\perp^2\right\}\right\rangle\left(z_{ob}\right)  
         - E\left\langle yr_\perp^2\right\rangle\left(z_{ob}\right),  
\label{pyaberration}
\end{eqnarray}
where, $\{\hat{A},\hat{B}\} = \hat{A}\hat{B} + \hat{B}\hat{A}$, the anticommutator of 
$\hat{A}$ and $\hat{B}$. With $g = g\left(z,z_{ob}\right)$, $h = h\left(z,z_{ob}\right)$, 
$g^\prime = g^\prime\left(z,z_{ob}\right)$, and $h^\prime = h^\prime\left(z,z_{ob}\right)$, 
the aberration coefficients are given by 
\begin{eqnarray}  
C & = &  \frac{1}{2}\int_{z_{ob}}^{z_{im}}dz\,\left\{\left(\alpha^4
       - \alpha\alpha^{\prime\prime}\right)h^4 + 2\alpha^2h^2{h^\prime}^2 
       + {h^\prime}^4\right\},  \nonumber \\ 
K & = &  \frac{1}{2}\int_{z_{ob}}^{z_{im}}dz\,\left\{\left(\alpha^4
       - \alpha\alpha^{\prime\prime}\right)gh^3 + \alpha^2(gh)^\prime hh^\prime 
       + g^\prime {h^\prime}^3\right\}, \nonumber \\ 
k & = &  \int_{z_{ob}}^{z_{im}}dz\,\left\{\left(\frac{1}{8}\alpha^{\prime\prime}
       - \frac{1}{2}\alpha^3\right)h^2 - \frac{1}{2}\alpha {h^\prime}^2\right\},  \nonumber \\ 
A & = &  \frac{1}{2}\int_{z_{ob}}^{z_{im}}dz\,\left\{\left(\alpha^4 
       - \alpha\alpha^{\prime\prime}\right)g^2h^2 + 2\alpha^2gg^\prime hh^\prime 
       + {g^\prime}^2{h^\prime}^2 - \alpha^2\right\},  \nonumber \\ 
a & = &  \int_{z_{ob}}^{z_{im}}dz\,\left\{\left(\frac{1}{4}\alpha^{\prime\prime} 
       - \alpha^3\right)gh - \alpha g^\prime h^\prime\right\},  \nonumber \\  
F & = &  \frac{1}{2}\int_{z_{ob}}^{z_{im}}dz\,\left\{\left(\alpha^4 
       - \alpha\alpha^{\prime\prime}\right)g^2h^2 + \alpha^2\left(g^2{h^\prime}^2
       + {g^\prime}^2h^2\right) + {g^\prime}^2{h^\prime}^2 + 2\alpha^2\right\},  \nonumber \\ 
D & = &  \frac{1}{2}\int_{z_{ob}}^{z_{im}}dz\,\left\{\left(\alpha^4 
       - \alpha\alpha^{\prime\prime}\right)g^3h + \alpha^2gg^\prime(gh)^\prime 
       + {g^\prime}^3h^\prime\right\},  \nonumber \\ 
d & = &  \int_{z_{ob}}^{z_{im}}dz\,\left\{\left(\frac{1}{8}\alpha^{\prime\prime}
       - \frac{1}{2}\alpha^3\right)g^2 - \frac{1}{2}\alpha {g^\prime}^2\right\}, \nonumber \\ 
E & = &  \frac{1}{2}\int_{z_{ob}}^{z_{im}}dz\,\left\{\left(\alpha^4 
       - \alpha\alpha^{\prime\prime}\right)g^4 + 2\alpha g^2{g^\prime}^2 + {g^\prime}^4\right\}.   
\label{aberexpns}
\end{eqnarray} 
In (\ref{xaberration}-\ref{yaberration}) $C$, $K$, $k$, $A$, $a$, $F$, $D$, and $d$ are, 
respectively, the coefficients of terms causing spherical aberration, isotropic coma, 
anisotropic coma, isotropic astigmatism, anisotropic astigmatism, field curvature, isotropic distortion, and anisotropic distortion in the image (see {\em e.g.}, \cite{Hawkes1989a} for 
detailed descriptions of these geometrical aberrations).  The terms with the coefficient $E$ 
in (\ref{pxaberration}-\ref{pyaberration}) cause aberrations of the ray gradients which do not 
affect the image by a single lens, but will have to be taken into account in multilens systems.  
This aberration has no name in electron optics and has been named pocus in light optics \cite{Dragt1986a, Khan2018b}.  The expressions for all the aberration coefficients in 
(\ref{aberexpns}) are the same as in classical electron optics.  

As noted earlier, a term like  
$\left\langle\hat{p}_x\hat{p}_\perp^2\right\rangle\left(z_{ob}\right)$ in 
$\langle\delta x\rangle\left(z_{ob}\right)$, for example, is  $\left(\langle\hat{p}_x\rangle\left(\langle\hat{p}_x\rangle^2 + 
\langle\hat{p}_y\rangle^2\right)\right)\left(z_{ob}\right)$ plus additional terms depending 
on quantum uncertainties which are uncontrollable in principle.  Thus, quantum uncertainties 
contribute to aberrations.  There are also other tiny quantum corrections to aberrations due 
to the $\hbar$-dedependent terms dropped from the quantum beam optical Hamiltonian.  The $\hbar$-dependent term $\propto r_\perp^2$ in $\hat{\mathcal{H}}_o^{(\hbar)}$ in 
(\ref{Hohbar}) is a paraxial term which, if added to $\hat{\mathcal{H}}_{o,p}$, will modify 
$\alpha(z)$ affecting the Larmor rotation, focal length, and the aberration coefficients, in 
a tiny way.  The other $\hbar$-dependent term in $\hat{\mathcal{H}}_o^{(\hbar)}$, 
$\propto r_\perp^4$, is a perturbation term which, if added to $\hat{\mathcal{H}}_o^\prime$, 
will modify the pocus and will have a tiny influence in a multilens system.   

Let us end this section with a note on the expression for the spherical aberration coefficient 
$C$ in (\ref{aberexpns}).  There are many expressions available for the various aberration 
coefficients in the literature.  As explained in \cite{Hawkes1989a}, though there may be several expressions for an aberration coefficient, very different in appearance, they are otherwise 
equivalent and can be obtained from one another by partial integration and replacing the second derivatives of $g$ and $h$ using the paraxial equation.  Thus, for $C$ there are various 
equivalent expressions, the most important and immensely influential being Scherzer's expression 
which showed that it is always positive for a round lens, free of space charge and with a static electromagnetic field, and hence the spherical aberration is unavoidable in such lenses.  We 
shall study the equivalence of the expression for $C$ in (\ref{aberexpns}) with Scherzer's 
expression and an expression used in \cite{Hawkes2002} (see also \cite{Hawkes1989a}).  The 
expressions in (\ref{aberexpns}) for all the aberration coefficients, obtained using quantum 
mechanics \cite{Jagannathan1996, Khan1997, Jagannathan2019} are the same as those obtained in 
\cite{Dragt1986b} using the Lie algebraic approach to classical electron optics which 
reproduces all the classical results exactly.  This is not surprising since the quantum electron 
beam optics developed in \cite{Jagannathan1996, Khan1997, Jagannathan2019} becomes the Lie 
algebraic approach to classical electron optics when the quantum $\longleftrightarrow$ classical correspondence rule (\ref{QCcorrespondence}) is used.  Lie algebraic methods have been developed 
extensively for classical charged particle beam optics, particularly accelerator beam optics 
(see {\em e.g.}, \cite{Dragt1986b, Dragt1988, Dragt2018, Radlicka2008}).  

Let us now start with the expression for $C$ given in (\ref{aberexpns}), rewritten as  
\begin{equation}  
C = \frac{1}{2}\int_{z_{ob}}^{z_{im}}dz\,\left\{\alpha^4h^4 + 2\alpha^2h^2{h^\prime}^2 
    - \alpha\alpha^{\prime\prime}h^4 + {h^\prime}^4\right\}.   
\label{a0}
\end{equation} 
Integrating the last term by parts, using the boundary conditions 
$h\left(z_{ob},z_{ob}\right) = 0$ and $h\left(z_{im},z_{ob}\right) = 0$ and the relation 
$h^{\prime\prime}\left(z,z_{ob}\right) = -\alpha(z)^2h\left(z,z_{ob}\right)$ following from 
the paraxial equation of motion (\ref{paraxialEq}) having $h\left(z,z_{ob}\right)$ as one of 
its solutions, we get 
\begin{equation}  
\int_{z_{ob}}^{z_{im}}dz\,{h^\prime}^4  
   = \int_{z_{ob}}^{z}dz\,h^\prime {h^\prime}^3 
   = h{h^\prime}^3\vert_{z_{ob}}^{z} - \int_{z_{ob}}^{z}dz\,3h{h^\prime}^2h^{\prime\prime} 
   = 3\int_{z_{ob}}^{z}dz\,\alpha^2h^2{h^\prime}^2.  
\label{a1}
\end{equation}   
Integrating the third term by parts leads to 
\begin{eqnarray}  
-\int_{z_{ob}}^{z_{im}}dz\,\alpha\alpha^{\prime\prime}h^4  
   & = & - \int_{z_{ob}}^{z_{im}}dz\,\alpha^{\prime\prime}\left(\alpha h^4\right)  \nonumber \\   
   & = & - \alpha^\prime\left(\alpha h^4\right)\vert_{z_{ob}}^{z_{im}} 
         + \int_{z_{ob}}^{z_{im}}dz\,\alpha^\prime\left(\alpha^\prime h^4 
         + 4 \alpha h^3h^\prime \right)  \nonumber \\  
   & = & \int_{z_{ob}}^{z_{im}}dz\,\left\{{\alpha^\prime}^2 h^4 
         + 4\alpha\alpha^\prime h^3h^\prime\right\}.  
\label{a2} 
\end{eqnarray} 
Integrating the term $\alpha\alpha^\prime h^3h^\prime$ by parts, and using the boundary 
conditons and the paraxial equation, we find 
\begin{eqnarray}  
\int_{z_{ob}}^{z_{im}}dz\,\alpha\alpha^\prime h^3h^\prime  
   & = & \int_{z_{ob}}^{z_{im}}dz\,\alpha^\prime\left(\alpha h^3h^\prime\right)  
     =   \alpha\left(\alpha h^3h^\prime\right)\vert_{z_{ob}}^{z_{im}}   
         - \int_{z_{ob}}^{z_{im}}dz\,\alpha\left(\alpha h^3h^\prime\right)^\prime  \nonumber \\ 
   & = & \int_{z_{ob}}^{z_{im}}dz\,\left\{-\alpha\alpha^\prime h^3h^\prime   
         - 3\alpha^2h^2{h^\prime}^2 + \alpha^4 h^4\right\}, 
\label{a3}
\end{eqnarray} 
which can be written in the equivalent form 
\begin{equation}  
\int_{z_{ob}}^{z_{im}}dz\,\alpha\alpha^\prime h^3h^\prime  
   = \frac{1}{2}\int_{z_{ob}}^{z_{im}}dz\,\left\{-3\alpha^2h^2{h^\prime}^2 + \alpha^4h^4\right\}.   
\label{a4}
\end{equation} 
Replacing the third and the last terms in the integral (\ref{a0}) using the relations 
(\ref{a1}), (\ref{a2}), and (\ref{a4}), we get    
\begin{equation}
C = \frac{1}{2}\int_{z_{ob}}^{z_{im}}dz\,\left\{5\alpha^2h^2{h^\prime}^2  
    + {\alpha^\prime}^2h^4 + 4\alpha\alpha^\prime h^3h^\prime + \alpha^4h^4\right\}.  
\end{equation}
Now, if we write the third term in the above equation as $(2+2)\alpha\alpha^\prime h^3h^\prime$ 
and replace one part $2\alpha\alpha^\prime h^3h^\prime$ using (\ref{a4}), keeping the other part $2\alpha\alpha^\prime h^3h^\prime$ as is, we obtain the celebrated result of Scherzer 
\begin{equation}  
C = \frac{1}{2}\int_{z_{ob}}^{z_{im}}dz\,\left\{2\alpha^4h^4 + h^2\left(h\alpha^{\prime} 
    + h^{\prime}\alpha\right)^2 + \alpha^2h^2{h^\prime}^2\right\}.   
\label{ScherzerC}
\end{equation} 
  
Now, we shall relate the expression for $C$ in (\ref{aberexpns}) to another expression 
\begin{equation}  
C = \frac{1}{48}\int_{z_{ob}}^{z_{im}}dz\,h^4\left\{4b^4 - bb^{\prime\prime} 
    + 5{b^\prime}^2\right\}, 
\label{HawkesC} 
\end{equation} 
in which 
\begin{equation}
b(z) = \sqrt{\frac{e}{2mU}}\,B(z) = 2\alpha(z), 
\label{alpha2b}
\end{equation} 
used in \cite{Hawkes2002} (see also \cite{Hawkes1989a}).  To this end we proceed as follows.  
As earlier, we use integration by parts, boundary conditions on $h$, and the paraxial equation.  
First, we find an expression for the integral of $\alpha\alpha^\prime h^3h^\prime$, different 
from (\ref{a3}), as given by  
\begin{eqnarray}  
\int_{z_{ob}}^{z_{im}}dz\,\alpha\alpha^\prime h^3h^\prime  
   & = & \left(\alpha\alpha^\prime h^3\right)h\vert_{z_{ob}}^{z_{im}}   
         - \int_{z_{ob}}^{z_{im}}dz\,h\left(\alpha\alpha^\prime h^3\right)^\prime  \nonumber \\ 
   & = & - \int_{z_{ob}}^{z_{im}}dz\,\left\{\left({\alpha^\prime}^2 
         + \alpha\alpha^{\prime\prime}\right)h^4 + 3\alpha\alpha^\prime h^3h^\prime\right\}, 
\label{b1}
\end{eqnarray} 
which can be written in the equivalent form 
\begin{equation}  
\int_{z_{ob}}^{z_{im}}dz\,\alpha\alpha^\prime h^3 h^\prime  
   = - \frac{1}{4}\int_{z_{ob}}^{z_{im}}dz\,\left({\alpha^\prime}^2
     + \alpha\alpha^{\prime\prime}\right)h^4.     
\label{b2}
\end{equation} 
Similarly, for the integral of $\alpha^2h^2{h^\prime}^2$ we find 
\begin{eqnarray}  
\int_{z_{ob}}^{z_{im}}dz\,\alpha^2h^2{h^\prime}^2  
   & = & \left(\alpha^2h^2h^\prime\right)h\vert_{z_{ob}}^{z_{im}}    
         - \int_{z_{ob}}^{z_{im}}dz\,h\left(\alpha^2h^2h^\prime\right)^\prime  \nonumber \\ 
   & = & \int_{z_{ob}}^{z_{im}}dz\,\left\{-2\alpha\alpha^\prime h^3h^\prime  
         - 2\alpha^2h^2{h^\prime}^2 + \alpha^4h^4\right\},    
\label{b3}
\end{eqnarray} 
which can be written in the equivalent form 
\begin{equation}  
\int_{z_{ob}}^{z_{im}}dz\,\alpha^2h^2{h^\prime}^2  
   = \frac{1}{3}\int_{z_{ob}}^{z}dz\,\left\{-2\alpha\alpha^\prime h^3h^\prime 
     + \alpha^4h^4\right\}.  
\label{b4}
\end{equation} 
Let us replace the last term in (\ref{a0}) using (\ref{a1}) leading to the expression 
\begin{equation}  
C = \frac{1}{2}\int_{z_{ob}}^{z_{im}}dz\,\left\{\alpha^4h^4 
    - \alpha\alpha^{\prime\prime}h^4 + 5\alpha^2h^2{h^\prime}^2\right\}.   
\label{b5}
\end{equation} 
Next, we replace the third term in (\ref{b5}) using (\ref{b4}) to obtain 
\begin{equation}  
C = \frac{1}{2}\int_{z_{ob}}^{z_{im}}dz\,\left\{\frac{8}{3}\alpha^4h^4 
    - \alpha\alpha^{\prime\prime}h^4 - \frac{10}{3}\alpha\alpha^\prime h^3h^\prime\right\}.   
\label{b6}
\end{equation} 
Then, we replace the last term in (\ref{b6}) using (\ref{b2}) to get (\ref{HawkesC}), 
\begin{eqnarray}  
C = \frac{1}{12}\int_{z_{ob}}^{z_{im}}dz\,h^4\left\{16\alpha^4 - \alpha\alpha^{\prime\prime} 
    + 5{\alpha^\prime}^2\right\}, 
\label{b7}
\end{eqnarray} 
which gives the expression for $C$ used in \cite{Hawkes2002} when $\alpha = b/2$ as seen in 
(\ref{alpha2b}).  

\section{Concluion} 
To summarize, we have considered the scalar theory of quantum mechanics of electron beam optics,  
at the single-particle level, derived from the Dirac equation using a Foldy-Wouthuysen-like transformation technique.  Guided by the Ehrenfest theorem, quantum averages of position and 
momentum of a beam electron in a plane transverse to the optic axis of an electron optical system 
have been identified with the classical ray coordinates.  Round magnetic electron lenses with 
Glaser and power law models for the axial magnetic field have been studied in particular.  We have found that in the paraxial approximation the quantum average of position obeys the classical 
paraxial equation of motion.  The fundamental solutions of the paraxial equations for the lenses considered, obtained by solving the differential equations, are well known.  In the case of the 
power law model lenses the fundamental solutions have also been constructed using the Peano-Baker series.  We have discussed the quantum propagator for paraxial propagation of the beam wave 
function along the optic axis of the system.  Quantum mechanics of aberrations due to deviation 
from paraxial condition has been discussed briefly.  Role of quantum uncertainties in the 
nonlinear part of the equations of motion for a  nonparaxial beam, and in aberrations, has been 
pointed out.  As remarked by Hawkes \cite{Hawkes2020}, quantum corrections to the classical theory 
of electron optics are `fortunately usually negligible', and a long article by Majert and Kohl \cite{Majert2019} on the simulation of atomically resolved elemental maps with a multislice 
algorithm for relativistic electrons shows that there are practical situations in which it is 
essential to use the Dirac theory.  Thus, with the development of nano-level technology in future electron beam optical devices quantum effects may have to be considered seriously.

\end{document}